\documentclass[aps,pra,english,twocolumn,showpacs,preprintnumbers,amsmath,amssymb,floatfix,superscriptaddress,longbibliography]{revtex4-1}

\usepackage[T1]{fontenc}
\usepackage[latin9]{inputenc}
\usepackage{graphicx}
\usepackage{amssymb}
\usepackage{babel}
\makeatletter
\usepackage{xspace}

\usepackage[version=3]{mhchem}

\usepackage{color}

\newcommand{\revisedtext}[1]{\textcolor{black}{#1}}

\makeatother
\usepackage{babel}
\makeatother

\begin{document}

\title{Axion Electrodynamics and the Casimir Effect}

\author{I. Brevik}
\email{iver.h.brevik@ntnu.no}
\affiliation{Department of Energy and Process Engineering, Norwegian University of Science and Technology, NO-7491 Trondheim, Norway}

\author{S. Pal}
\affiliation{Centre of Excellence ENSEMBLE3 Sp. z o. o., Wolczynska Str. 133, 01-919, Warsaw, Poland}

\author{Y. Li}
\affiliation{Department of Physics, Nanchang University, Nanchang 330031, China}
\affiliation{Institute of Space Science and Technology, Nanchang University, Nanchang 330031, China}

\author{A. Gholamhosseinian}
\affiliation{Department of Physics, Ferdowsi University of Mashhad, Mashad, Iran}

\author{M. Bostr{\"o}m}
\affiliation{Centre of Excellence ENSEMBLE3 Sp. z o. o., Wolczynska Str. 133, 01-919, Warsaw, Poland}
\affiliation{Chemical and Biological Systems Simulation Lab, Centre of New Technologies, University of Warsaw, Banacha 2C, 02-097 Warsaw, Poland}

\date{\today}%

\begin{abstract}
We present a concise review of selected parts of axion electrodynamics and its application to Casimir physics.  We present the general formalism including the boundary conditions at a dielectric surface, derive the dispersion relation in the case where the axion parameter has a constant spatial derivative in the direction normal to the conducting plates, and calculate the Casimir energy for the simple case of scalar electrodynamics using dimensional regularization.
\end{abstract}

\maketitle


%
\section{Introduction}
\label{Intro}
\par The axion concept has actually a long history. It was introduced by Peccei and Quinn back in 1977~\cite{peccei77, peccei08} in connection with the CP problem in high-energy physics.
But later it has been understood as related to a natural extra term in the electromagnetic Lagrangian, with a direct formal connection to materials like topological insulators and thus of obvious practical interest. So, we first describe the basic properties of topological materials.
Topological insulators (TIs) are the new phases of matter, which exhibit unique electronic properties due to their nontrivial topological characteristics, and were discovered in 2005~\cite{kane2005z,kane2005quantum}. These materials have insulating states inside the bulk with a bulk energy gap separating the highest occupied electronic band from the lowest empty band, like an ordinary insulator, while conducting states exist on their surfaces in the case of three-dimensional TIs or edges in the case of two-dimensional (2D) TIs, which are topologically protected (robust to local defects, imperfections, and disorders), by time-reversal symmetry~\cite{hasan2010colloquium}. In 2006, the TI phase was theoretically predicted~\cite{bernevig2006quantum} and experimentally realized in a CdTe-HgTe-CdTe quantum well: this quantum well behaves in bulk as an insulator. However, the electric current was observed across the interface, i.e., it behaves like a conductor in the surface region~\cite{qi2010quantum}. Additionally, we know from band theory that conductors do not have a gap between their valence and conduction bands. In contrast, insulators are defined as materials with a gap between them. The most notable aspect is that Maxwell's equations are unable to explain the experimental behaviors of topological insulators.  Notably, this kind of behavior was previously suggested by Frank Wilczek~\cite{wilczek1987two} in 1987, along with the possibility that it could be described by the axion electrodynamics he~\cite{wilczek1978problem} and Steven Weinberg~\cite{weinberg1978new} developed. Their initial aim was to explain the breaking of combined symmetries of charge conjugation and parity in strong interactions.

\par With a topological invariant called the $\mathbb{Z}_2$ invariant, we can distinguish trivial insulators from topological insulators. Topological materials have interesting features, and one of them is the magnetoelectric effect caused by a term called the $\theta$ term. Since this term, referred to as the magnetoelectric polarizability, has exactly the same form as the action describing the coupling between a photon and an axion, these magnetoelectric phenomena are often depicted with the axion electrodynamics.
In the presence of time-reversal symmetry, $\theta$ takes on a quantized value $\theta=\pi ({\rm mod}\ 2\pi)$ for topological insulators, and $\theta = 0$ for ordinary insulators~\cite{fu2007topological,liu2011quantum,fukui2005chern,sekine2021axion}. The value of $\theta$, nevertheless, can be arbitrary in systems with broken time-reversal symmetry, even depending on space and time as $\theta(r,t)$, like in various semi-metalic phases. Moreover, when the dynamics of the axion field is included, the existence of new quasiparticles, such as the axion polariton~\cite{li2010dynamical}, was also proposed. For more details, please refer to Ref.~\cite{sekine2021axion} and the references therein.

\par In a separate contribution to the special issue, we reviewed the semi-classical electrodynamics and its link to Casimir physics.
{Notably, the history of this remarkable effect dates back to 1948 when it was predicted by Hendrik Casimir~\cite{Casi,CasimirPolder48}. A formidable, and highly effective, theory for the retarded dispersion force between a pair of planar surfaces interacting across an intervening media was developed in the 1950s by Lifshitz and co-workers\,\cite{Lifshitz,Dzya}.
From the late 1960s groups from around the world explored if a theory based upon the classical Maxwell's equations combined with the Planck quantization of light could by itself lead to a simple and useful semi-classical theory for van der Waals, Lifshitz, and Casimir interactions\,\cite{ParNin1969,NinhamParsegianWeiss1970}. As one important example: a semi-classical derivation of Casimir effects in magnetic media was presented by Peter Richmond and Barry Ninham\,\cite{Richmond_1971} already in 1971.
 Much theoretical\,\cite{Rich71,Rich73,Bost2000,Bord,EPJDNinham2014,Maofeng2014PhysRevB.89.201407,Estesodoi:10.1021/jp511851z,PhysRevB.97.125421,Estesosdoi:10.1021/acs.jpclett.9b02030,Esteso4layerPCCP2020,Ninham_Brevik_Bostrom_2022,LiBrevikMalyiBostromPhysRevE.108.034801,BostromRizwanHarshanBrevikLiPerssonMalyi2023spontaneous,MostKlim2023}
 and experimental works\,\cite{Haux,AndSab,Lamo1997,HarrisPhysRevA.62.052109,DeccaPhysRevLett.91.050402,Feiler2008,Munday2009,SomersGarrettPalmMunday_CasimirTorque} have followed the last 50 years.

 More information related to Casimir effects in traditional systems can be found in the extensive literature\,\cite{Maha,milton01,Pars,Bordagbook,Ser,Buhmann12a,Ser2018}. A lot of works have been carried out with a focus on predictive theories, nanobiotechnological applications, and novel materials growth and characterization. Notably, going beyond the standard applications, specific ion effects in both biological systems and colloid chemistry have been proposed to occur partly due to ionic dispersion potentials\,\cite{Bost2001} acting on ions in salt solutions\,\cite{ParsonsNinham2009,ParsonsNinham2012,Medda2012doi:10.1021/la3035984,DuignanNinhamParsons2013solvation,ParsonsSalis2015,Bostr2016,ThiyamFiedlerBuhmannPerssonBrevikBostromParsons2018}.
This leads to a non-linear coupling of electrodynamical and electrostatic interactions with a proposed role behind the so-called (ion-specific) Hofmeister effect\,\cite{NinhamYaminsky1997}.
Most interestingly, Casimir and van der Waals interactions may also have an impact on the growth of ice clusters within mist\,\cite{BostromvdWicegrowthmistPCCP2023} and clouds\,\cite{LUENGOMARQUEZMacDowell2021,LuengoMarquez_IzquierdoRuiz_MacDowell2022}.
It has been, and still is, relevant to explore the limits of validity of the different theories for dispersion forces (e.g. between two layered surfaces). }
A most natural extension of this conventional semi-classical electrodynamics, as well as quantum electrodynamics, in media is to allow for an extra pseudoscalar field, called conventionally the axion field  $a(x)$ ($x$ means here spacetime), pervading in the whole volume. Some of the pioneering papers on axion electrodynamics are listed in Refs.~\cite{peccei77,peccei08,sikivie83,weinberg78,preskill83,abbott83,dine83,sikivie08,sikivie14}. More recent works can be found in Refs.~\cite{millar17,liu22,li91,lawson19,qingdong19,sikivie03,mcdonald20,zyla20,arza20,carenza20,leroy20,oullet19,arza19,
qiu17,dror21,fukushima19,tobar19,bae22,adshead20,tobar22,derocco18,brevik22,brevik22a,brevik23,favitta23}.
We will present an easy-to-follow and concise review of the current understanding of axion electrodynamics from our point of view.

\section{Axion Electrodynamics}

\label{secintro}

\par

\revisedtext{Different methods have been suggested to investigate the electromagnetic characteristics of 3D magnetic topological insulators, see for example Refs.\,\cite{sikivie83,weinberg78}, which is based on axion field theory ~\cite{qi2011topological,qi2008topological}. This approach introduces an additional term to the conventional Maxwell electromagnetic action, expressed as follows
\begin{equation}
S_A = \frac{e^2}{32\pi^2\hbar c} \int d^3\Vec{r}\, dt\, \theta\, \varepsilon^{\mu\nu\alpha\beta}F_{\mu\nu}F_{\alpha\beta},
\label{a}
\end{equation}
where $\varepsilon^{mnab}$ represents the completely antisymmetric tensor, $F_{mn}$ is the Maxwell field strength tensor, and $q$ denotes the axion coupling strength. Originally proposed in the context of quantum chromodynamics (QCD) to address the strong CP problem ~\cite{peccei1977cp,peccei1977constraints}, the axion is a hypothetical pseudoscalar particle and has also been considered a potential candidate for cosmological dark matter. Since Equation (\ref{a}) bears mathematical similarities to the description of cosmological/QCD axions, the term "axion" is used in the context of topological insulators. However, the axial coupling in topological insulators is related to the presence of a surface quantum Hall effect~\cite{lu2021casimir}.
}

\par The reason why $a(x)$ is a pseudoscalar quantity is that the mentioned extra term in the Lagrangian implies a two-photon interaction with the electromagnetic field, and the pseudoscalar property of $a(x)$ ensures that its product with the polar vector ${\bf E}$ together with the axial vector ${\bf B}$ becomes a scalar in the Lagrangian. We shall in the following highlight some essential properties of the axion formalism, assuming a dielectric environment with the permittivity $\varepsilon$ and the permeability $\mu$ being constants. It means that the constitutive relations are simply  ${\bf D}=\varepsilon {\bf E}, \, {\bf B}=\mu {\bf H}$. The presentation in the following is largely based upon our earlier papers \cite{brevik22,brevik22a,brevik23,favitta23}. When magnetoelectric effects occur in topological material, the magnetic induction ${\bf B}$ changes the electric displacement vector ${\bf D}$, and the magnetic field intensity ${\bf H}$ is in turn influenced by the electric field. Then, the relations for ${\bf D}$ and ${\bf B}$ should be changed, ${\bf D}=\varepsilon {\bf E}-\frac{\theta \alpha}{\pi} {\bf B}$, ${\bf H}=\frac{\bf B}{\mu}+\frac{\theta \alpha}{\pi} {\bf E}$, where $\alpha$ is  the fine-structure constant.
{ These constitutive relations were given  in \cite{martin2019magnetoelectric}  and in references therein; cf. also the later Ref.~\cite{nogueira2022absence}.}
 \revisedtext{As a result, these two polarizations are coupled, meaning that the electromagnetic boundary conditions are off-diagonal components of the Fresnel coefficients\,\cite{woods2020perspective}.}

{ In relation to our previous comments in Sec. 1, it is important to note parallelism between two analogous yet distinct phenomena. In the following discussion, our focus will be on the axion approach rooted in the Peccei-Quinn formalism, while a formal analogy arises with polariton excitations in condensed media. Our coupling constant $g_{a\gamma\gamma}$ below will thus refer to the axion case. One might be interested in the corresponding coupling constant in the polarization case also, but this is a complicated subject into which we will not enter here. Interested readers may consult for instance, the paper~\cite{to2022strong}, to get a detailed information about a strong coupling between a topological insulator and a III-V heterostructure.   }

 \bigskip

\noindent{\it Basic equations.~}
We choose the metric convention with signature $g_{00}= -1$, and introduce two field tensors $F_{\alpha\beta}$ and $H^{\alpha\beta}$, with $\alpha,\beta$ running from 0 to 3. The components of the original field tensor $F_{\alpha\beta}$ are as in vacuum, $F_{0i}=-E_i, F_{jk}=\epsilon_{ijk}B_i$ for $i,j,k=1,2,3$, while the components of the contravariant response tensor $H^{\alpha\beta}$ are  $ H^{0i}=D_i, H_{jk}=\epsilon_{ijk}H_i$.

Including the  pseudoscalar axion field $a= a(x)$,   we get for the  Lagrangian
 \begin{equation}
{\cal{L}}= -\frac{1}{4}F_{\alpha\beta}{H}^{\alpha\beta} +{\bf A \cdot J}-\rho \Phi -\frac{1}{2} \partial_\mu a\partial^\mu a-\frac{1}{2}m_a^2a^2       - \frac{1}{4}\theta(x) F_{\alpha\beta}\tilde{F}^{\alpha\beta}, \label{1}
\end{equation}
in which $\tilde{F}^{\alpha\beta}=\epsilon^{\alpha\beta\gamma\delta}F_{\gamma\delta}/2$, and we have defined the nondimensional field amplitude as
\begin{equation}
\theta(x)= g_{a\gamma\gamma}a(x).
\end{equation}
\revisedtext{The constant for the coupling of axion with two photons can be defined as follows,}
\begin{equation}
g_{a\gamma\gamma}= g_\gamma \frac{\alpha}{\pi}\frac{1}{f_a},
\end{equation}
\revisedtext{where $g_\gamma$ is a constant depending on the specific model used,} 
\revisedtext{is usually taken to be} $0.36$~\cite{sikivie03}. 
\revisedtext{ The $f_a$ and $\alpha$ represented the fine structure and the axion decay constant respectively, and for $f_a$, it is commonly assumed that its value is on the order of $10^{12}~$GeV.} 
\revisedtext{
The Lagrangian's last term (\ref{1}), ${\cal{L}}_{a\gamma\gamma}$, can be written as ${\cal{L}}_{a\gamma\gamma} =  \theta(x){\bf E\cdot B}$, explicitly showing the pseudoscalar property of $\theta(x)$.}

\par 
\revisedtext{The expression (\ref{1}) can be used to obtain the generalized Maxwell equations}
\begin{equation}
{\bf \nabla \cdot D}= \rho-{\bf B\cdot \nabla}\theta, \label{5}
\end{equation}
\begin{equation}
{\bf \nabla \times H}= {\bf J}+\frac{\partial}{\partial t}{\bf D}+\frac{\partial\theta}{\partial t}{\bf B}+{\bf \nabla}\theta\times {\bf E}, \label{7}
\end{equation}
\begin{equation}
{\bf \nabla \cdot B}=0, \label{8}
\end{equation}
\begin{equation}
{\bf \nabla \times E} = -\frac{\partial}{\partial t}{\bf B}. \label{9}
\end{equation}
\revisedtext{These equations are general, and there are no restrictions on the spacetime variation of $a(x)$. The equations are relativistically covariant.} 
\revisedtext{It is crucial that the constitutive relations stated previously maintain their simple form ${\bf D}=\varepsilon {\bf E}, \, {\bf B}=\mu {\bf H}$ only in the rest system, and the electromagnetic formalism's covariance is achieved by introducing {\it two} distinct field tensors, $F_{\mu\nu}$ and $H^{\mu\nu}$.} 
\revisedtext{The field equations describing the system are}
\begin{equation}
\nabla^2 {\bf E}-\varepsilon\mu \frac{\partial^2}{\partial t^2}{\bf E}=    {\bf \nabla (\nabla \cdot E)}
 +\mu \frac{\partial}{\partial t}{\bf J}+ \mu \frac{\partial}{\partial t}\left[\frac{\partial\theta}{\partial t}{\bf B}+ {\bf \nabla}\theta{\bf \times E}\right], \label{10}
\end{equation}
\begin{equation}
\nabla^2 {\bf H}-\varepsilon\mu \frac{\partial^2}{\partial t^2}{\bf H}= -{\bf \nabla \times J}-{\bf \nabla \times }[\frac{\partial\theta}{\partial t}{\bf B}+{\bf \nabla}\theta{\bf \times E}]. \label{11}
\end{equation}
In practice, the influence from axions is typically small in our case here. We do not consider the field equations for the axions explicitly for simplicity and clarity.

\par The field equations above contain the second-order derivatives of $\theta$. These may conveniently be removed, if we assume there is a strong external magnetic field ${\bf B}_e = B_e\hat{ \bf z}$ presents. Then assuming the axion field
\begin{equation}
a(t)= a_0\cos \omega_at,
\end{equation}
we can separate out the part ${\bf E}_a(t)$ related to  the $\ddot{\theta}$ term. From the governing equation for ${\bf E}_a(t)$, ignoring the current ${\bf J}$ as the axion-related field,
\begin{equation}
\nabla^2 {\bf E}_a-\varepsilon\mu \frac{\partial^2}{\partial t^2}{\bf E}_a=    \mu \ddot{\theta}{\bf B}_e .\label{ 14a}
\end{equation}
Then,
\begin{equation}
{\bf E}_a(t)= -\frac{1}{\varepsilon}E_0\cos \omega_at \,\hat{\bf z},
\end{equation}
where
\begin{equation}
E_0 =\theta_0 B_e.
\end{equation}
After this separation, the field equations  take the reduced forms
 \begin{equation}
\nabla^2 {\bf E}-\varepsilon\mu \frac{\partial^2}{\partial t^2}{\bf E}=  {\bf \nabla (\nabla \cdot E)} +\mu \frac{\partial}{\partial t}{ \bf J}+  \mu [ \dot{\theta}\frac{\partial}{\partial t}{\bf B} + {\bf \nabla}\theta{\bf \times \frac{\partial}{\partial t}{E}}], \label{12}
\end{equation}
\begin{equation}
\nabla^2 {\bf H}-\varepsilon\mu \frac{\partial^2}{\partial t^2}{\bf H} = -{\bf \nabla \times J}-\left[   \dot{\theta}{\bf \nabla \times B}
 + ({\bf \nabla}\theta){\bf \nabla \cdot E}
 -({\bf \nabla}\theta \cdot {\bf \nabla})  {\bf E}      \right]. \label{13}
\end{equation}
{ In the following we will allow $\theta$ to be spatially varying, but will neglect the second order derivatives, i.e., $ \partial_i\partial_j \theta \approx 0.$ It means, the model excludes situations where spatial boundaries lead to $\delta$ and  $\delta'$ type terms. Actually, as shown in Eq.~(\ref{betadefinition}) below, we will in the mathematical analysis take $\theta$ to vary linearly in space over the field region of interest. There will accordingly be no second-order terms in the field region, while the electromagnetic boundary conditions must be imposed at the boundaries. }

{ It is to be noted that the equations contain the dynamical fields $\bf E$ and $\bf B$ only. }

\bigskip

\noindent {\it Hybrid form of Maxwell's equations. Boundary conditions.~} It turns out that one can construct a hybrid form of Maxwell's equations from which the generalized boundary conditions at a dielectric surface follow in a very transparent way.
Introduce new fields ${\bf D}_\gamma$ and ${\bf H}_\gamma$ via
\begin{equation}
\left(\begin{array}{ll}
{\bf D}_\gamma \\
{\bf H}_\gamma
\end{array}
\right)=
\left( \begin{array}{ll}
\varepsilon  & \theta \\
-\theta      & 1/\mu
\end{array}
\right)
\left(\begin{array}{ll}
{\bf E} \\
{\bf B}
\end{array}
\right), \label{constitutiverelation}
\end{equation}
which shows  how ${\bf D}_\gamma, {\bf H}_\gamma$  relate to the response tensor $H^{\mu\nu}$.
The hybrid Maxwell equations thus become similar to those  in usual electrodynamics,
\begin{equation}
{ \bf \nabla \times H}_\gamma = {\bf J}+ \frac{\partial}{\partial t}{\bf D}_\gamma, \quad {\bf \nabla \cdot D}_\gamma = \rho,
\end{equation}
\begin{equation}
  { \bf \nabla \times E} = - \frac{\partial}{\partial t}{\bf  B},  \quad {\bf \nabla \cdot B} =0.
\end{equation}
The boundary conditions at a dielectric boundary are then immediate,
\begin{equation}
E_\perp = E_{\gamma,\perp} + E_{a,\perp} \quad \rm{ is ~continuous},
\end{equation}

\begin{equation}
E_\perp,H_{\gamma,\perp} \quad \rm{are ~continuous},
\end{equation}
\begin{equation}
D_{\gamma,\parallel},B_{\parallel} \quad {\rm are ~continuous},
\end{equation}
(the symbol $\perp$ and $\parallel$ mean perpendicular and parallel to the normal, thus parallel and perpendicular to the surface).  An important quantity is the property of Poynting's vector, ${\bf S= E\times H}$.
Taking the $z$ component $S_z$ orthogonal to a dielectric surface, we see that
\begin{equation}
S_{1z}=S_{2z},
\end{equation}
showing that the energy flux density is continuous across the surface dividing media 1 and 2. This is as one would expect as the surface is {\it at rest}; the force acting on it is not able to do any mechanical work.

\bigskip

\noindent {\it Dispersion  relations.~}  Use now  of the standard plane wave expansion
\begin{equation}
{\bf E} \propto e^{i({\bf k\cdot x}-\omega t)}, \quad  {\bf  k}= (k_x,k_y,k_z). \label{expansion}
\end{equation}
We start from the reduced Maxwell equations (\ref{12}) and (\ref{13}), and restrict ourselves to the case where $a(x)$ to vary with space only,
\begin{equation}
 \boldmath{\beta}= {\bf \nabla}\theta, \label{betadefinition}
\end{equation}
with $\bm{\beta}$ assumed  constant.
{ This form is mathematically convenient and often used in the literature. The form is of the same kind as assumed for Weyl semimetals ~\cite{kargarian2015theory,wilson2015repulsive} where the gradient of the axion is related to the separation of Weyl nodes in the Brillouin zone. One may here note that the case of topological insulators is different, as in such a case the the gradient of the axion is taken to be zero except at the interfaces between trivial and nontrivial phases.  The situation is however complex, and it should be mentioned that the configuration given here is very close to the one reported in Ref.~\cite{martin2016green}, where a topological insulator slab is placed between two perfect conducting plates.  We also point out that very useful papers on the new material as those showing exotic Hall effects, as those on TRS-broken semimetals, Chern insulators, etc, can be found in the References~\cite{woods2016materials,khusnutdinov2019casimir,woods2020perspective}. }

Starting  from Eqs.~(\ref{12}) and (\ref{13}), assuming $\rho={\bf J} =0$, we find the determinental equation determining the dispersion relations. There are two dispersive branches. The first is a "normal" one, satisfying
\begin{equation}
\varepsilon\mu \omega^2 = {\bf k}^2, \label{firstbranch}
\end{equation}
corresponding to waves independent of the axions and polarizing parallel to $\bm{\beta}$. The second branch, polarizing perpendicular to $\bm{\beta}$, should have
\begin{equation}
\varepsilon\mu \omega^2= {\bf k}^2 \pm \mu \beta \omega. \label{alpha}
\end{equation}
showing the splitting of this branch into two modes, equally separated from the normal mode on both sides. This sort of splitting has been encountered before under various circumstances; cf., for instance, Refs.~\cite{sikivie03,mcdonald20,brevik21}.

\par The following property of this kind of material should be noticed: the dispersive property does not influence the electromagnetic energy density. In a complex representation, the energy density can be written as
\begin{equation}
W_{\rm disp}= \frac{1}{4}\left[\frac{d(\varepsilon_{\rm eff}\omega)}{d\omega}|{\bf E}|^2+|{\bf H}|^2\right],
\end{equation}
where we have assumed for a moment that   the medium is nonmagnetic, and  we have introduced an effective permittivity $\varepsilon_{\rm eff}$ such that
\begin{equation}
{\bf k}^2 = \varepsilon_{\rm eff}(\omega) \omega^2. \label{dispersion}
\end{equation}
It then follows that
\begin{equation}
\frac{d(\varepsilon_{\rm eff}\omega)}{d\omega} = \varepsilon,
\end{equation}
which shows that the dispersive correction disappears. The energy density is the same as if dispersion were absent. This property is not quite trivial.

\bigskip

\noindent {\it Dispersion  relations, when $\theta$ is time dependent.~} We introduce the variable $\alpha$ as
\begin{equation}
\alpha = \dot{\theta}
\end{equation}
and assume $\alpha$ constant. The dispersion takes the form
\begin{equation}
\varepsilon\mu\omega^2 = {\bf k}^2 \pm \mu\alpha |{\bf k}|. \label{alfabet}
\end{equation}
We may here for a nonmagnetic medium introduce the refractive index,  $n_{\rm eff}=\sqrt{\varepsilon_{\rm eff}}$,  and so get
\begin{equation}
n_{\rm eff}(\omega)= \sqrt{\varepsilon +\frac{\alpha^2}{4\omega^2}} \pm \frac{\alpha}{2\omega}.
\end{equation}
Given the assumed smallness of the axion contributions,  we restrict the parameter values to the region $(\beta + \alpha)/k_z \ll 1$.

If $k_z$ is the wave number for photon in the $z$ direction, corresponding to  axions with mass  $m_a \sim \omega=10^{-5}~$eV, we  have $k_z=10^{-5}~$eV, so that the condition above can be rewritten as
\begin{equation}\label{betacond}
\beta+\alpha \ll 10^{-5}~{\rm eV} \,\left(\frac{m_a}{10^{-5}~{\rm eV}}\right).
\end{equation}

\section{Energy-momentum considerations}
\par It is interesting to consider the electromagnetic energy-momentum tensor, in interaction with the axion field. As this is a nonclosed physical system, the four-force density will in general be different from zero. The system is in this way analogous to the electromagnetic field in a medium in ordinary electrodynamics, since also, in that case, the system is nonlosed because the influence from the mechanical system itself is not accounted for (this is the point giving rise to the classic Abraham-Minkowski problem). We assume in this section that $\varepsilon$ and $\mu$ can vary in space and time, but do not restrict the derivatives with respect to space or time to be constants.

 We may start with the electromagnetic energy density,
\begin{equation}
W = \frac{1}{2}( \bf {E\cdot D} + {\bf H\cdot B} ). \label{17}
\end{equation}
Together with the Poynting vector,
\begin{equation}
{\bf S}= {\bf E\times H}, \label{15}
\end{equation}
 this leads to the energy conservation equation
\begin{equation}
{\bf \nabla \cdot S} + \dot{W}= -{\bf E \cdot J}-\dot{\theta}{ \bf (E\cdot B)}. \label{16}
\end{equation}
 There is thus an exchange of electromagnetic energy with the axion "medium" if $\bf E$ and $\bf B$ are different from zero and $\theta(t)$ is time-varying, even if ${\bf J}=0.$

 A more delicate question is the expression for the momentum density $\bf g$. In accordance with
 Planck's principle of inertia of energy, ${\bf g=S}/c^2$, one would expect the Abraham momentum density $  {\bf g}^{\rm A}$    to be right,
\begin{equation}
{\bf g}^{\rm A}= {\bf E\times H}. \label{18}
\end{equation}
We notice that  the  Maxwell stress tensor,
\begin{equation}
T_{ik}= E_iD_k+H_iB_k-\frac{1}{2}\delta_{ik}({\bf E\cdot D+H\cdot B}), \label{19}
\end{equation}
 is common for the Abraham and Minkowski alternatives, $T_{ik}^{\rm A} = T_{ik}^{\rm M} \equiv T_{ik}$.
The momentum conservation equation can thus be expressed in the form
\begin{equation}
\partial_kT_{ik}-\dot{g}_i^{\rm A}= f_i^{\rm A},
\end{equation}
where $f_i^{\rm A}$ are the components of Abraham's force density
\begin{equation}
{\bf f}^{\rm A}= \rho {\bf E} + ({\bf J\times B}) + (\varepsilon \mu -1)\frac{\partial}{\partial t}({\bf E\times H})
-g_{a\gamma\gamma} {\bf (E\cdot B)\nabla}a.         \label{20}
\end{equation}
The third term on the right hand side,  the Abraham term,    has experimentally turned up only in a few experiments, mainly at low frequencies where the mechanical oscillations of a test body are directly detectable.

   In optics,  the Abraham force will fluctuate out. It is therefore mathematically simpler, and in accordance with all observational experience in optics, to include the Abraham momentum (physically, a mechanical accompanying momentum) in the effective field momentum. Therewith, the momentum density becomes simply the Minkowski momentum ${\bf g}^{\rm M}$, given by
\begin{equation}
{\bf g}^{\rm M}= {\bf D\times B}. \label{21}
\end{equation}
Momentum conservation equation  in the Minkowski case yields
\begin{equation}
\partial_kT_{ik}-\frac{\partial}{\partial t}g_i^{\rm M}= f_i^{\rm M},
\end{equation}
where
\begin{equation}
{\bf f}^{\rm M}= \rho {\bf E}+({\bf J\times B})
-  {\bf (E\cdot B)} {\nabla\theta}.        \label{22}
\end{equation}
Consider the relativistically covariant form for the energy-momentum balance:  Minkowski's energy-momentum tensor
\begin{equation}
S_\mu^{{\rm M}\nu}  = F_{\mu\alpha}H^{\nu\alpha}-\frac{1}{4}g_\mu^\nu F_{\alpha\beta}H^{\alpha\beta}, \label{23}
\end{equation}
 has the same form in all inertial frames. The conservation equations for electromagnetic energy and momentum can  be written as
\begin{equation}
-\partial_\nu S_\mu^{{\rm M}\nu}  = f_\mu ^{\rm M}, \label{24}
\end{equation}
where $f_\mu^{\rm M} = (f_0, {\bf f}^{\rm M})$ is the four-force density. In the rest system,
\begin{equation}
f_0= {\bf E}\cdot {\bf J}+ {\bf (E\cdot B)}\dot{\theta}, \label{25}
\end{equation}
where  $f_0^{\rm M}= f_0^{\rm A} \equiv f_0$.

\section{Casimir effect between two plates}

 We will here restrict ourselves to the simplest case, namely scalar electrodynamics, meaning that the photons' vector property are included but not their spin. We assume $\alpha =0$, and also zero temperature. Assume the usual setup where there are two large metallic plates, with a gap $L$, orthogonal to the $z$ direction.  In the intervening region there is an axion field $a(z)$ whose amplitude increases linearly with $z$,
\begin{equation}
a(z)= \frac{a_0z}{L}= \beta z, \quad 0<z<L,
\end{equation}
where $a_0$, the amplitude at $z=L$, is fixed. There is no external magnetic field. {The magnitude of $\beta$ is not restricted to be small.  The smallness expansion in the present theory applies rather to the spatial variation of the axion field, as embodied in the restriction $\partial_i\partial_j \theta \approx 0$, as also mentioned above.}

We mention that the reduced Maxwell equations in this case can be written as
\begin{equation}
\nabla^2 {\bf E}-\varepsilon\mu \ddot{\bf E}= g_{a\gamma\gamma}\mu \beta {\bf \hat{z}\times \dot{E}}, \label{23}
\end{equation}
\begin{equation}
\nabla^2 {\bf H}-\varepsilon\mu \ddot{\bf H} = g_{a\gamma\gamma} \beta \partial_z\bf E. \label{24}
\end{equation}
We get two dispersive branches, as before. The first, corresponding to Eq.~(\ref{firstbranch}), can now be written
\begin{equation}
|{\bf k}| = \sqrt{\varepsilon \mu}\,\omega, \quad k_z = \frac{\pi n}{L}, \quad n=1,2,3...,
\end{equation}
showing no dependence on the axions. The second branch can be written as
\begin{equation}
 \varepsilon \mu \omega^2={\bf k}^2\pm g_{a\gamma\gamma} \beta \omega,
 \label{29}
\end{equation}
which can be recast in the form
\begin{equation}
\omega = \frac{1}{\sqrt{\varepsilon \mu}}\left[ |{\bf k}| \pm \frac{1}{2} g_{a\gamma\gamma} \beta \sqrt{\frac{\mu}{\varepsilon}}\right],\label{30}
\end{equation}
in view of the smallness of $g_{a\gamma \gamma}^2$. This branch, composed of two modes, lies very close to the first mode above. For a given $\omega$ there are in all three different values of $|{\bf k}|$.

\par Consider now  the zero-temperature zero-point energy $\cal{E}$ of the field, defined by
\begin{equation}
{\cal E}= \frac{1}{2}\sum \omega.
\end{equation}
From the second branch (\ref{30}) only, we get, per unit plate area,
\begin{equation}
{\cal{E}}= \frac{1}{2\sqrt{\varepsilon\mu}}\sum_{n=1}^\infty\left[
\int \frac{d^2 k_\perp}{(2\pi)^2}
  \sqrt{k_\perp^2+\frac{\pi^2n^2}{L^2}}
 \pm \frac{ g_{a\gamma\gamma}}{2L^2}   \sqrt{\frac{\mu}{\varepsilon}} \right], \label{31}
\end{equation}
where ${\bf k}_\perp$ is the component of $\bf k$ orthogonal to the surface normal \footnote{Note that the present definition of $\beta$ differs from that in Ref.~\cite{brevik21}.}. This expression can be further treated  using dimensional regularization (cf., for instance, Ref.~\cite{milton01}), with the result \cite{brevik21}
\begin{equation}
{\cal{E}} = \frac{1}{\sqrt{\varepsilon\mu}}\left[ -\frac{\pi^2}{1440}\frac{1}{L^3} \mp  \frac {g_{a\gamma\gamma} \beta} {4L^2}  \sqrt{ \frac{\mu}{\varepsilon}} \right].  \label{37}
\end{equation}
Of main interest is the contribution from the photons and axions moving in the $z$ direction. We associate this with the Casimir energy ${\cal{E}}_C$. Mathematically, the derivation involves the Hurwitz zeta function.  We obtain
\begin{equation}
{\cal{E}}_C = \frac{1}{4\sqrt{\varepsilon\mu}}\left( -\frac{\pi}{6L} \pm \frac{1}{2}g_{a\gamma\gamma}\beta \sqrt{\frac{\mu}{\varepsilon}}\right) \frac{1}{L^2}.        \label{44}
\end{equation}
Here the first term comes from scalar photons propagating in the $z$ direction, while the second term is the axionic contribution. With respect to the inverse $L$ dependence, the Casimir energies for the electrodynamic and the axion parts behave similarly as one would expect.
In the above, the upper and lower signs match. In Eq.~(\ref{44}), the small axion-induced increase of the Casimir energy arises from the superluminal mode in the dispersion relation (\ref{30})  (meaning that the group velocity is larger than $1/\sqrt{\varepsilon\mu}$). This mode corresponds to a weak repulsive Casimir force. The other mode corresponds to a weak attractive force.

\section{An axion echo from reflection in outer space}

Assume that the axion field is of the form   $a=a(t) = a_0\sin \omega_at$, as usually done for the outer space, and assume  $\varepsilon = \mu = 1$. In view of the smallness of the axion velocity,  $v \sim   10^{-3}c$, we have approximately  $\omega_a =m_a$. An estimated value of $m_a = 10~\mu$eV is common.
If the axions are associated with dark matter, we have to do with a very low energy density, $\rho_{\rm DM}= 0.45~$GeV/cm$^3$\,\cite{liu22}.
 We will adopt the simple form
\begin{equation}
\rho_a= \frac{1}{2}m_a^2a_0^2. \label{57}
\end{equation}
Assume now that an electromagnetic beam is sent from the Earth to an axion cloud. We take the initial form of the beam to be Gaussian,
 \begin{equation}
 {\bf A}_0(x,t)=  c\,e^{-x^2/2D^2}\cos k_0x \,\hat{\bf y}, \label{58}
 \end{equation}
 with ${\bf A}_0$ as the vector potential, and the constant $D$  as the Gaussian width. Making a Fourier decomposition,
  \begin{equation}
 {\bf A}_0(x.t)= \frac{1}{2} \int_{-\infty}^\infty \left[{\bf A}_0(k)e^{i(kx-\omega t)} + {\bf A}_0^* (k)e^{-i(kx-\omega t)} \right] dk, \label{59}
 \end{equation}
 we have as inversion
 \begin{equation}
 {\bf A}_0(k)= \frac{1}{2\pi}\int_{-\infty}^\infty e^{-ikx}\left[ {\bf A}_0(x,0)+\frac{i}{\omega}\frac{\partial {\bf A}_0}{\partial t}(x,0)\right]dx. \label{60}
 \end{equation}
 Then imposing the condition  ${\partial {\bf A}_0/\partial t}(x,0)=0$, we  get
 \begin{equation}
 {\bf A}_0(k)= {\bf A}_0^+(k)+ {\bf A}_0^- (k), \label{62}
 \end{equation}
 where
 \begin{equation}
 {\bf A}_0^+(k)= \frac{cD}{2\sqrt{2\pi}}\exp\left[ -\frac{1}{2}D^2(k-k_0)^2\right]\, \hat{\bf y}. \label{62}
 \end{equation}
The expression for ${\bf A}_0^-(k)$ implies that  $k-k_0 \rightarrow k+k_0$. The right-moving wave, to be considered henceforth, is
\begin{equation}
{\bf A}_0^+(x,t)= \int_0^\infty {\bf A}_0^+(k)\cos(kx-\omega t)dk,\quad \omega = k >0. \label{63}
\end{equation}
We will omit  the superscript. The  incident wave fields are
${\bf E}_0(x,t)= -{\dot{\bf A}}_ 0(x,t),
 \quad {\bf H}_0(x,t)= {\bf \nabla \times A}_0(x,t). $

The incident wave will interact with the axion cloud.  As $\rho = {\bf J}=0$ we  see from (\ref{20}) or (\ref{22}) that ${\bf f}=0$, in accordance with the assumed homogeneity   of the cloud.
Also, the component $f_0$ in (\ref{25}) is zero, because  ${\bf E}_0$ and ${\bf H}_0$ are orthogonal.
 We therefore we go back to the field equations, from which  we get
\begin{equation}
\nabla^2{\bf A}-\ddot{\bf A}= -g_{a\gamma\gamma}\dot{a}{\bf \nabla \times A}. \label{65}
\end{equation}
We here let ${\bf A} \rightarrow {\bf A}_0$ on the RHS. Neglecting $\nabla^2{\bf A}$ on the left, and using Eqs.~(\ref{63}) we obtain
\begin{equation}
\ddot{\bf A}(x,t)=      -g_{a\gamma\gamma}a_0\omega_a\int_0^\infty A_0(k)k\sin(kx-\omega t)\cos \omega_at\,dk \, \hat{\bf z}, \label{66}
\end{equation}
still with $\omega= k$.

In complex notation, extracting the resonance term,
\begin{equation}
\ddot{\bf A}(x,t)= \frac{1}{2}g_{a\gamma\gamma}a_0\omega_a{\rm Im}\int_0^\infty A_0(k)ke^{i(kx-\omega t+\omega_at)}\, \hat{\bf z}. \label{67}
\end{equation}
We construct a new quantity  ${\bf A}(k,t)$ as
\begin{equation}
{\bf A}(x,t)= {\rm Im}\int_0^\infty e^{ikx}{\bf A}(k,t)dk, \label{67}
\end{equation}
and can thus write
\begin{equation}
\ddot{\bf A}(k,t)=-\frac{1}{2}g_{a\gamma\gamma}a_0\omega_aA_0(k)ke^{-i(\omega-\omega_a)t}\, \hat{\bf z}. \label{68}
\end{equation}
With yet another quantity
\begin{equation}
{\bf {\cal{A}}}(k,t)={\bf A}(k,t)e^{-i\omega t}, \label{69}
\end{equation}
we have
\begin{equation}
\ddot{\bf A}(k,t)= \ddot{{\bf {\cal{A}}}}(k,t) e^{i\omega t} +2i\omega \dot{{\bf {\cal{A}}}}(k,t) e^{i\omega t}- \omega^2 {\bf {\cal{A}}}(k,t) e^{i\omega t}, \label{70}
\end{equation}
Now keeping the resonance term only, we get
\begin{equation}
\dot{{\bf {\cal{A}}}}(k,t)  = \frac{i}{4}g_{a\gamma\gamma}a_0\omega_aA_0(k)e^{-i(2\omega-\omega_a)t}\, \hat{\bf z}. \label{71}
\end{equation}
We integrate over  $t$,
\begin{equation}
{\bf {\cal{A}}}(k,t) = -g_{a\gamma\gamma}a_0\omega_a\frac{cD}{8\sqrt{2\pi}}\exp\left[ -\frac{1}{2}D^2(k-k_0)^2\right]\frac{e^{-i(2\omega-\omega_a)t}}{2\omega-\omega_a}\, \hat{\bf z}. \label{72}
\end{equation}
The resonance at $\omega= \omega/2$ is as expected:  an incoming photon $2\omega$ can split an axion into two components with the same mass,  $m_a=\omega_a$. We focus on  the imaginary part of the above expression, and use the relation
\begin{equation}
\lim \frac{\sin \alpha x}{\pi x}\big|_{\alpha \rightarrow \infty}= \delta(x), \label{73}
\end{equation}
to get
\begin{equation}
{\rm{ Im}{{{\bf {\cal{A}}}}}}(k,t)  = g_{a\gamma\gamma}a_0\omega_a\frac{cD}{16}\sqrt{\frac{\pi}{2}}\exp\left[ -\frac{1}{2}D^2(k-k_0)^2\right]\delta(\omega-\frac{1}{2}\omega_a)\, \hat{\bf z}. \label{74}
\end{equation}
From this, the axion echo can be found. This sort of calculation was pioneered by Sikivie {\it et al.}
\cite{sikivie03} and \cite{arza19}. The present generalized form, containing the Gaussian width $D$,
was given by Brevik {\it et al.} \cite{brevik22}.

\section{Discussions and Future Outlooks}

We would like to conclude with the following brief points:

\noindent 1. As already mentioned, the influence of axions, at least in cosmology, is expected to be very weak. The cosmological axion energy density is often expected to be about $0.40~$GeV/cm$^3$, corresponding to an axion mass of about $10^{-5}~$eV and a relative velocity of about $10^{-3}$. Various experiments and proposals of experiments have been launched:

 (a)  The haloscope experiment, proposed by   Sikivie {\it et al.} \cite{sikivie03}), in which the aim is to detect resonances between the electromagnetic eigenfrequencies of a dielectric cylinder and the axions (cf. also Refs.~\cite{millar17,brevik22}. So far, no such resonance has been detected.

 (b) The idea, also due to Sikivie {\it et al.} \cite{sikivie03}), to observe the axions via their electromagnetic "echo"  returned back to the Earth from an outer cloud (cf. also Ref.~\cite{brevik22}).

 (c) The broadband solenoidal haloscope proposed by Liu et al. \cite{liu22}, which proposes to make use of the axion "antenna" effect to focus the electromagnetic radiation emitted from dielectric boundaries towards a detector.

\noindent 2. The above treatment provides a general review of axion electrodynamics and is, in principle, not limited to the semi-classical case. This constraint applies similarly to ordinary electrodynamics, usually when distances are small or temperatures are high.

\noindent 3. The axion formalism is useful as regards application to topological insulators. Thus the constitutive relations (\ref{constitutiverelation}) can formally be taken over to this kind of modern material science as they stand. The case of chiral materials, for instance, a Faraday material, is more complicated since the coupling parameter $\theta$ becomes imaginary. Cf., for instance, Ref.~\cite{qingdong19}.

\noindent 4. {To conclude: we have in the current contribution to the special issue on "75 Years of the Casimir Effect: Advances and Prospects", presented a concise summary of the basics of axion electrodynamics linking it to the general field of Casimir physics. Notably, additional contributions to the Casimir interaction are observed that occur as direct consequences of the extra pseudoscalar axion field. Novel physics, based on improved materials characterization requiring new and improved physical models, is likely to be discovered in the years to come.}

\begin{acknowledgments}
 This research is part of the project No. 2022/47/P/ST3/01236 co-funded by the National Science Centre and the European Union's Horizon 2020 research and innovation programme under the Marie Sk{\l}odowska-Curie grant agreement No. 945339. We also thank the "ENSEMBLE3 - Centre of Excellence for nanophotonics, advanced materials and novel crystal growth-based technologies" project (GA No. MAB/2020/14) carried out within the International Research Agendas programme of the Foundation for Polish Science co-financed by the European Union under the European Regional Development Fund, the European Union's Horizon 2020 research and innovation programme Teaming for Excellence (GA. No. 857543) for support of this work.
\end{acknowledgments}

\bibliography{main}

\begin{thebibliography}{111}%
\makeatletter
\providecommand \@ifxundefined [1]{%
 \@ifx{#1\undefined}
}%
\providecommand \@ifnum [1]{%
 \ifnum #1\expandafter \@firstoftwo
 \else \expandafter \@secondoftwo
 \fi
}%
\providecommand \@ifx [1]{%
 \ifx #1\expandafter \@firstoftwo
 \else \expandafter \@secondoftwo
 \fi
}%
\providecommand \natexlab [1]{#1}%
\providecommand \enquote  [1]{``#1''}%
\providecommand \bibnamefont  [1]{#1}%
\providecommand \bibfnamefont [1]{#1}%
\providecommand \citenamefont [1]{#1}%
\providecommand \href@noop [0]{\@secondoftwo}%
\providecommand \href [0]{\begingroup \@sanitize@url \@href}%
\providecommand \@href[1]{\@@startlink{#1}\@@href}%
\providecommand \@@href[1]{\endgroup#1\@@endlink}%
\providecommand \@sanitize@url [0]{\catcode `\\12\catcode `\$12\catcode
  `\&12\catcode `\#12\catcode `\^12\catcode `\_12\catcode `\%12\relax}%
\providecommand \@@startlink[1]{}%
\providecommand \@@endlink[0]{}%
\providecommand \url  [0]{\begingroup\@sanitize@url \@url }%
\providecommand \@url [1]{\endgroup\@href {#1}{\urlprefix }}%
\providecommand \urlprefix  [0]{URL }%
\providecommand \Eprint [0]{\href }%
\providecommand \doibase [0]{http://dx.doi.org/}%
\providecommand \selectlanguage [0]{\@gobble}%
\providecommand \bibinfo  [0]{\@secondoftwo}%
\providecommand \bibfield  [0]{\@secondoftwo}%
\providecommand \translation [1]{[#1]}%
\providecommand \BibitemOpen [0]{}%
\providecommand \bibitemStop [0]{}%
\providecommand \bibitemNoStop [0]{.\EOS\space}%
\providecommand \EOS [0]{\spacefactor3000\relax}%
\providecommand \BibitemShut  [1]{\csname bibitem#1\endcsname}%
\let\auto@bib@innerbib\@empty
\bibitem [{\citenamefont {Peccei}\ and\ \citenamefont
  {Quinn}(1977{\natexlab{a}})}]{peccei77}%
  \BibitemOpen
  \bibfield  {author} {\bibinfo {author} {\bibfnamefont {R.~D.}\ \bibnamefont
  {Peccei}}\ and\ \bibinfo {author} {\bibfnamefont {Helen~R.}\ \bibnamefont
  {Quinn}},\ }\bibfield  {title} {\enquote {\bibinfo {title} {$\mathrm{CP}$
  conservation in the presence of pseudoparticles},}\ }\href {\doibase
  10.1103/PhysRevLett.38.1440} {\bibfield  {journal} {\bibinfo  {journal}
  {Phys. Rev. Lett.}\ }\textbf {\bibinfo {volume} {38}},\ \bibinfo {pages}
  {1440--1443} (\bibinfo {year} {1977}{\natexlab{a}})}\BibitemShut {NoStop}%
\bibitem [{\citenamefont {Peccei}(2008)}]{peccei08}%
  \BibitemOpen
  \bibfield  {author} {\bibinfo {author} {\bibfnamefont {R.~D.}\ \bibnamefont
  {Peccei}},\ }\href {\doibase 10.1007/978-3-540-73518-2} {\emph {\bibinfo
  {title} {Axions: Theory, Cosmology, and Experimental Searches (Editors M.
  Kuster and G. Raffelt and B. Beltr{\'a}n)}}},\ edited by\ \bibinfo {editor}
  {\bibfnamefont {M.}~\bibnamefont {Kuster}}, \bibinfo {editor} {\bibfnamefont
  {G.}~\bibnamefont {Raffelt}}, \ and\ \bibinfo {editor} {\bibfnamefont
  {B.}~\bibnamefont {Beltr{\'a}n}}\ (\bibinfo  {publisher} {Springer Berlin,
  Heidelberg},\ \bibinfo {year} {2008})\BibitemShut {NoStop}%
\bibitem [{\citenamefont {Kane}\ and\ \citenamefont
  {Mele}(2005{\natexlab{a}})}]{kane2005z}%
  \BibitemOpen
  \bibfield  {author} {\bibinfo {author} {\bibfnamefont {Charles~L}\
  \bibnamefont {Kane}}\ and\ \bibinfo {author} {\bibfnamefont {Eugene~J}\
  \bibnamefont {Mele}},\ }\bibfield  {title} {\enquote {\bibinfo {title} {Z 2
  topological order and the quantum spin hall effect},}\ }\href {\doibase
  10.1103/PhysRevLett.95.146802} {\bibfield  {journal} {\bibinfo  {journal}
  {Physical review letters}\ }\textbf {\bibinfo {volume} {95}},\ \bibinfo
  {pages} {146802} (\bibinfo {year} {2005}{\natexlab{a}})}\BibitemShut
  {NoStop}%
\bibitem [{\citenamefont {Kane}\ and\ \citenamefont
  {Mele}(2005{\natexlab{b}})}]{kane2005quantum}%
  \BibitemOpen
  \bibfield  {author} {\bibinfo {author} {\bibfnamefont {Charles~L}\
  \bibnamefont {Kane}}\ and\ \bibinfo {author} {\bibfnamefont {Eugene~J}\
  \bibnamefont {Mele}},\ }\bibfield  {title} {\enquote {\bibinfo {title}
  {Quantum spin hall effect in graphene},}\ }\href {\doibase
  10.1103/PhysRevLett.95.226801} {\bibfield  {journal} {\bibinfo  {journal}
  {Physical review letters}\ }\textbf {\bibinfo {volume} {95}},\ \bibinfo
  {pages} {226801} (\bibinfo {year} {2005}{\natexlab{b}})}\BibitemShut
  {NoStop}%
\bibitem [{\citenamefont {Hasan}\ and\ \citenamefont
  {Kane}(2010)}]{hasan2010colloquium}%
  \BibitemOpen
  \bibfield  {author} {\bibinfo {author} {\bibfnamefont {M~Zahid}\ \bibnamefont
  {Hasan}}\ and\ \bibinfo {author} {\bibfnamefont {Charles~L}\ \bibnamefont
  {Kane}},\ }\bibfield  {title} {\enquote {\bibinfo {title} {Colloquium:
  topological insulators},}\ }\href {\doibase 10.1103/revmodphys.82.3045}
  {\bibfield  {journal} {\bibinfo  {journal} {Reviews of modern physics}\
  }\textbf {\bibinfo {volume} {82}},\ \bibinfo {pages} {3045} (\bibinfo {year}
  {2010})}\BibitemShut {NoStop}%
\bibitem [{\citenamefont {Bernevig}\ \emph {et~al.}(2006)\citenamefont
  {Bernevig}, \citenamefont {Hughes},\ and\ \citenamefont
  {Zhang}}]{bernevig2006quantum}%
  \BibitemOpen
  \bibfield  {author} {\bibinfo {author} {\bibfnamefont {B~Andrei}\
  \bibnamefont {Bernevig}}, \bibinfo {author} {\bibfnamefont {Taylor~L}\
  \bibnamefont {Hughes}}, \ and\ \bibinfo {author} {\bibfnamefont {Shou-Cheng}\
  \bibnamefont {Zhang}},\ }\bibfield  {title} {\enquote {\bibinfo {title}
  {Quantum spin hall effect and topological phase transition in hgte quantum
  wells},}\ }\href {\doibase 10.1126/science.1133734} {\bibfield  {journal}
  {\bibinfo  {journal} {science}\ }\textbf {\bibinfo {volume} {314}},\ \bibinfo
  {pages} {1757--1761} (\bibinfo {year} {2006})}\BibitemShut {NoStop}%
\bibitem [{\citenamefont {Qi}\ and\ \citenamefont
  {Zhang}(2010)}]{qi2010quantum}%
  \BibitemOpen
  \bibfield  {author} {\bibinfo {author} {\bibfnamefont {Xiao-Liang}\
  \bibnamefont {Qi}}\ and\ \bibinfo {author} {\bibfnamefont {Shou-Cheng}\
  \bibnamefont {Zhang}},\ }\bibfield  {title} {\enquote {\bibinfo {title} {The
  quantum spin hall effect and topological insulators},}\ }\href {\doibase
  10.1063/1.3293411} {\bibfield  {journal} {\bibinfo  {journal} {Physics
  Today}\ }\textbf {\bibinfo {volume} {63}},\ \bibinfo {pages} {33--38}
  (\bibinfo {year} {2010})}\BibitemShut {NoStop}%
\bibitem [{\citenamefont {Wilczek}(1987)}]{wilczek1987two}%
  \BibitemOpen
  \bibfield  {author} {\bibinfo {author} {\bibfnamefont {Frank}\ \bibnamefont
  {Wilczek}},\ }\bibfield  {title} {\enquote {\bibinfo {title} {Two
  applications of axion electrodynamics},}\ }\href {\doibase
  10.1103/PhysRevLett.58.1799} {\bibfield  {journal} {\bibinfo  {journal}
  {Physical review letters}\ }\textbf {\bibinfo {volume} {58}},\ \bibinfo
  {pages} {1799} (\bibinfo {year} {1987})}\BibitemShut {NoStop}%
\bibitem [{\citenamefont {Wilczek}(1978)}]{wilczek1978problem}%
  \BibitemOpen
  \bibfield  {author} {\bibinfo {author} {\bibfnamefont {Frank}\ \bibnamefont
  {Wilczek}},\ }\bibfield  {title} {\enquote {\bibinfo {title} {Problem of
  strong p and t invariance in the presence of instantons},}\ }\href {\doibase
  10.1103/PhysRevLett.40.279} {\bibfield  {journal} {\bibinfo  {journal}
  {Physical Review Letters}\ }\textbf {\bibinfo {volume} {40}},\ \bibinfo
  {pages} {279} (\bibinfo {year} {1978})}\BibitemShut {NoStop}%
\bibitem [{\citenamefont {Weinberg}(1978{\natexlab{a}})}]{weinberg1978new}%
  \BibitemOpen
  \bibfield  {author} {\bibinfo {author} {\bibfnamefont {Steven}\ \bibnamefont
  {Weinberg}},\ }\bibfield  {title} {\enquote {\bibinfo {title} {A new light
  boson?}}\ }\href {\doibase 10.1103/PhysRevLett.40.223} {\bibfield  {journal}
  {\bibinfo  {journal} {Physical Review Letters}\ }\textbf {\bibinfo {volume}
  {40}},\ \bibinfo {pages} {223} (\bibinfo {year}
  {1978}{\natexlab{a}})}\BibitemShut {NoStop}%
\bibitem [{\citenamefont {Fu}\ and\ \citenamefont
  {Kane}(2007)}]{fu2007topological}%
  \BibitemOpen
  \bibfield  {author} {\bibinfo {author} {\bibfnamefont {Liang}\ \bibnamefont
  {Fu}}\ and\ \bibinfo {author} {\bibfnamefont {Charles~L}\ \bibnamefont
  {Kane}},\ }\bibfield  {title} {\enquote {\bibinfo {title} {Topological
  insulators with inversion symmetry},}\ }\href {\doibase
  10.1103/PhysRevB.76.045302} {\bibfield  {journal} {\bibinfo  {journal}
  {Physical Review B}\ }\textbf {\bibinfo {volume} {76}},\ \bibinfo {pages}
  {045302} (\bibinfo {year} {2007})}\BibitemShut {NoStop}%
\bibitem [{\citenamefont {Liu}\ \emph {et~al.}(2011)\citenamefont {Liu},
  \citenamefont {Feng},\ and\ \citenamefont {Yao}}]{liu2011quantum}%
  \BibitemOpen
  \bibfield  {author} {\bibinfo {author} {\bibfnamefont {Cheng-Cheng}\
  \bibnamefont {Liu}}, \bibinfo {author} {\bibfnamefont {Wanxiang}\
  \bibnamefont {Feng}}, \ and\ \bibinfo {author} {\bibfnamefont {Yugui}\
  \bibnamefont {Yao}},\ }\bibfield  {title} {\enquote {\bibinfo {title}
  {Quantum spin hall effect in silicene and two-dimensional germanium},}\
  }\href {\doibase 10.1103/PhysRevLett.107.076802} {\bibfield  {journal}
  {\bibinfo  {journal} {Physical review letters}\ }\textbf {\bibinfo {volume}
  {107}},\ \bibinfo {pages} {076802} (\bibinfo {year} {2011})}\BibitemShut
  {NoStop}%
\bibitem [{\citenamefont {Fukui}\ \emph {et~al.}(2005)\citenamefont {Fukui},
  \citenamefont {Hatsugai},\ and\ \citenamefont {Suzuki}}]{fukui2005chern}%
  \BibitemOpen
  \bibfield  {author} {\bibinfo {author} {\bibfnamefont {Takahiro}\
  \bibnamefont {Fukui}}, \bibinfo {author} {\bibfnamefont {Yasuhiro}\
  \bibnamefont {Hatsugai}}, \ and\ \bibinfo {author} {\bibfnamefont {Hiroshi}\
  \bibnamefont {Suzuki}},\ }\bibfield  {title} {\enquote {\bibinfo {title}
  {Chern numbers in discretized brillouin zone: efficient method of computing
  (spin) hall conductances},}\ }\href {\doibase 10.1143/JPSJ.74.1674}
  {\bibfield  {journal} {\bibinfo  {journal} {Journal of the Physical Society
  of Japan}\ }\textbf {\bibinfo {volume} {74}},\ \bibinfo {pages} {1674--1677}
  (\bibinfo {year} {2005})}\BibitemShut {NoStop}%
\bibitem [{\citenamefont {Sekine}\ and\ \citenamefont
  {Nomura}(2021)}]{sekine2021axion}%
  \BibitemOpen
  \bibfield  {author} {\bibinfo {author} {\bibfnamefont {Akihiko}\ \bibnamefont
  {Sekine}}\ and\ \bibinfo {author} {\bibfnamefont {Kentaro}\ \bibnamefont
  {Nomura}},\ }\bibfield  {title} {\enquote {\bibinfo {title} {Axion
  electrodynamics in topological materials},}\ }\href {\doibase
  10.1063/5.0038804} {\bibfield  {journal} {\bibinfo  {journal} {Journal of
  Applied Physics}\ }\textbf {\bibinfo {volume} {129}} (\bibinfo {year}
  {2021}),\ 10.1063/5.0038804}\BibitemShut {NoStop}%
\bibitem [{\citenamefont {Li}\ \emph {et~al.}(2010)\citenamefont {Li},
  \citenamefont {Wang}, \citenamefont {Qi},\ and\ \citenamefont
  {Zhang}}]{li2010dynamical}%
  \BibitemOpen
  \bibfield  {author} {\bibinfo {author} {\bibfnamefont {R.~D.}\ \bibnamefont
  {Li}}, \bibinfo {author} {\bibfnamefont {J.}~\bibnamefont {Wang}}, \bibinfo
  {author} {\bibfnamefont {X.~L.}\ \bibnamefont {Qi}}, \ and\ \bibinfo {author}
  {\bibfnamefont {S.~C.}\ \bibnamefont {Zhang}},\ }\bibfield  {title} {\enquote
  {\bibinfo {title} {Dynamical axion field in topological magnetic
  insulators},}\ }\href@noop {} {\bibfield  {journal} {\bibinfo  {journal}
  {Nature Phys.}\ }\textbf {\bibinfo {volume} {6}},\ \bibinfo {pages}
  {284--288} (\bibinfo {year} {2010})}\BibitemShut {NoStop}%
\bibitem [{\citenamefont {Casimir}(1948)}]{Casi}%
  \BibitemOpen
  \bibfield  {author} {\bibinfo {author} {\bibfnamefont {H.~B.~G.}\
  \bibnamefont {Casimir}},\ }\bibfield  {title} {\enquote {\bibinfo {title} {On
  the attraction between two perfectly conducting plates},}\ }\href@noop {}
  {\bibfield  {journal} {\bibinfo  {journal} {K. Ned. Akad. Wet.}\ }\textbf
  {\bibinfo {volume} {51}},\ \bibinfo {pages} {793} (\bibinfo {year}
  {1948})}\BibitemShut {NoStop}%
\bibitem [{\citenamefont {Casimir}\ and\ \citenamefont
  {Polder}(1948)}]{CasimirPolder48}%
  \BibitemOpen
  \bibfield  {author} {\bibinfo {author} {\bibfnamefont {H.~B.~G.}\
  \bibnamefont {Casimir}}\ and\ \bibinfo {author} {\bibfnamefont
  {D.}~\bibnamefont {Polder}},\ }\bibfield  {title} {\enquote {\bibinfo {title}
  {{The influence of retardation on the London--van der Waals forces}},}\
  }\href {\doibase 10.1103/PhysRev.73.360} {\bibfield  {journal} {\bibinfo
  {journal} {Phys. Rev.}\ }\textbf {\bibinfo {volume} {73}},\ \bibinfo {pages}
  {360--372} (\bibinfo {year} {1948})}\BibitemShut {NoStop}%
\bibitem [{\citenamefont {Lifshitz}(1956)}]{Lifshitz}%
  \BibitemOpen
  \bibfield  {author} {\bibinfo {author} {\bibfnamefont {E.~M.}\ \bibnamefont
  {Lifshitz}},\ }\bibfield  {title} {\enquote {\bibinfo {title} {The theory of
  molecular attractive forces between solids},}\ }\href@noop {} {\bibfield
  {journal} {\bibinfo  {journal} {Sov. Phys. JETP}\ }\textbf {\bibinfo {volume}
  {2}},\ \bibinfo {pages} {73--83} (\bibinfo {year} {1956})}\BibitemShut
  {NoStop}%
\bibitem [{\citenamefont {Dzyaloshinskii}\ \emph {et~al.}(1961)\citenamefont
  {Dzyaloshinskii}, \citenamefont {Lifshitz},\ and\ \citenamefont
  {Pitaevskii}}]{Dzya}%
  \BibitemOpen
  \bibfield  {author} {\bibinfo {author} {\bibfnamefont {I.E.}\ \bibnamefont
  {Dzyaloshinskii}}, \bibinfo {author} {\bibfnamefont {E.M.}\ \bibnamefont
  {Lifshitz}}, \ and\ \bibinfo {author} {\bibfnamefont {L.P.}\ \bibnamefont
  {Pitaevskii}},\ }\bibfield  {title} {\enquote {\bibinfo {title} {{The general
  theory of van der Waals forces}},}\ }\href {\doibase
  10.1080/00018736100101281} {\bibfield  {journal} {\bibinfo  {journal} {Adv.
  Phys.}\ }\textbf {\bibinfo {volume} {10}},\ \bibinfo {pages} {165--209}
  (\bibinfo {year} {1961})}\BibitemShut {NoStop}%
\bibitem [{\citenamefont {Parsegian}\ and\ \citenamefont
  {Ninham}(1969)}]{ParNin1969}%
  \BibitemOpen
  \bibfield  {author} {\bibinfo {author} {\bibfnamefont {VA}~\bibnamefont
  {Parsegian}}\ and\ \bibinfo {author} {\bibfnamefont {BW}~\bibnamefont
  {Ninham}},\ }\bibfield  {title} {\enquote {\bibinfo {title} {Application of
  the lifshitz theory to the calculation of van der waals forces across thin
  lipid films},}\ }\href {\doibase 10.1038/2241197a0} {\bibfield  {journal}
  {\bibinfo  {journal} {Nature}\ }\textbf {\bibinfo {volume} {224}},\ \bibinfo
  {pages} {1197--1198} (\bibinfo {year} {1969})}\BibitemShut {NoStop}%
\bibitem [{\citenamefont {Ninham}\ \emph {et~al.}(1970)\citenamefont {Ninham},
  \citenamefont {Parsegian},\ and\ \citenamefont
  {Weiss}}]{NinhamParsegianWeiss1970}%
  \BibitemOpen
  \bibfield  {author} {\bibinfo {author} {\bibfnamefont {B.~W.}\ \bibnamefont
  {Ninham}}, \bibinfo {author} {\bibfnamefont {V.~A.}\ \bibnamefont
  {Parsegian}}, \ and\ \bibinfo {author} {\bibfnamefont {G.~H.}\ \bibnamefont
  {Weiss}},\ }\bibfield  {title} {\enquote {\bibinfo {title} {{On the
  macroscopic theory of temperature dependent Van der Waals forces}},}\ }\href
  {\doibase 10.1007/BF01020441} {\bibfield  {journal} {\bibinfo  {journal} {J.
  Stat. Phys.}\ }\textbf {\bibinfo {volume} {2}},\ \bibinfo {pages} {323}
  (\bibinfo {year} {1970})}\BibitemShut {NoStop}%
\bibitem [{\citenamefont {Richmond}\ and\ \citenamefont
  {Ninham}(1971{\natexlab{a}})}]{Richmond_1971}%
  \BibitemOpen
  \bibfield  {author} {\bibinfo {author} {\bibfnamefont {P}~\bibnamefont
  {Richmond}}\ and\ \bibinfo {author} {\bibfnamefont {B~W}\ \bibnamefont
  {Ninham}},\ }\bibfield  {title} {\enquote {\bibinfo {title} {A note on the
  extension of the lifshitz theory of van der waals forces to magnetic
  media},}\ }\href {\doibase 10.1088/0022-3719/4/14/014} {\bibfield  {journal}
  {\bibinfo  {journal} {Journal of Physics C: Solid State Physics}\ }\textbf
  {\bibinfo {volume} {4}},\ \bibinfo {pages} {1988--1993} (\bibinfo {year}
  {1971}{\natexlab{a}})}\BibitemShut {NoStop}%
\bibitem [{\citenamefont {Richmond}\ and\ \citenamefont
  {Ninham}(1971{\natexlab{b}})}]{Rich71}%
  \BibitemOpen
  \bibfield  {author} {\bibinfo {author} {\bibfnamefont {P.}~\bibnamefont
  {Richmond}}\ and\ \bibinfo {author} {\bibfnamefont {B.W.}\ \bibnamefont
  {Ninham}},\ }\bibfield  {title} {\enquote {\bibinfo {title} {{Calculations,
  using Lifshitz theory, of the height vs. thickness for vertical liquid helium
  films}},}\ }\href {\doibase https://doi.org/10.1016/0038-1098(71)90459-5}
  {\bibfield  {journal} {\bibinfo  {journal} {Solid State Comm.}\ }\textbf
  {\bibinfo {volume} {9}},\ \bibinfo {pages} {1045 -- 1047} (\bibinfo {year}
  {1971}{\natexlab{b}})}\BibitemShut {NoStop}%
\bibitem [{\citenamefont {Richmond}\ \emph {et~al.}(1973)\citenamefont
  {Richmond}, \citenamefont {Ninham},\ and\ \citenamefont {Ottewill}}]{Rich73}%
  \BibitemOpen
  \bibfield  {author} {\bibinfo {author} {\bibfnamefont {P}~\bibnamefont
  {Richmond}}, \bibinfo {author} {\bibfnamefont {B.W}\ \bibnamefont {Ninham}},
  \ and\ \bibinfo {author} {\bibfnamefont {R.H}\ \bibnamefont {Ottewill}},\
  }\bibfield  {title} {\enquote {\bibinfo {title} {{A theoretical study of
  hydrocarbon adsorption on water surfaces using Lifshitz theory}},}\ }\href
  {\doibase https://doi.org/10.1016/0021-9797(73)90243-9} {\bibfield  {journal}
  {\bibinfo  {journal} {J. Coll. Interf. Sci.}\ }\textbf {\bibinfo {volume}
  {45}},\ \bibinfo {pages} {69 -- 80} (\bibinfo {year} {1973})}\BibitemShut
  {NoStop}%
\bibitem [{\citenamefont {Bostr\"om}\ and\ \citenamefont
  {Sernelius}(2000)}]{Bost2000}%
  \BibitemOpen
  \bibfield  {author} {\bibinfo {author} {\bibfnamefont {M.}~\bibnamefont
  {Bostr\"om}}\ and\ \bibinfo {author} {\bibfnamefont {Bo~E.}\ \bibnamefont
  {Sernelius}},\ }\bibfield  {title} {\enquote {\bibinfo {title} {{Thermal
  Effects on the Casimir Force in the 0.1-5\,$\mu$\,m Range}},}\ }\href
  {\doibase 10.1103/PhysRevLett.84.4757} {\bibfield  {journal} {\bibinfo
  {journal} {Phys. Rev. Lett.}\ }\textbf {\bibinfo {volume} {84}},\ \bibinfo
  {pages} {4757} (\bibinfo {year} {2000})}\BibitemShut {NoStop}%
\bibitem [{\citenamefont {Bordag}\ \emph {et~al.}(2000)\citenamefont {Bordag},
  \citenamefont {Geyer}, \citenamefont {Klimchitskaya},\ and\ \citenamefont
  {Mostepanenko}}]{Bord}%
  \BibitemOpen
  \bibfield  {author} {\bibinfo {author} {\bibfnamefont {M.}~\bibnamefont
  {Bordag}}, \bibinfo {author} {\bibfnamefont {B.}~\bibnamefont {Geyer}},
  \bibinfo {author} {\bibfnamefont {G.~L.}\ \bibnamefont {Klimchitskaya}}, \
  and\ \bibinfo {author} {\bibfnamefont {V.~M.}\ \bibnamefont {Mostepanenko}},\
  }\bibfield  {title} {\enquote {\bibinfo {title} {{Casimir Force at Both
  Nonzero Temperature and Finite Conductivity}},}\ }\href {\doibase
  10.1103/PhysRevLett.85.503} {\bibfield  {journal} {\bibinfo  {journal} {Phys.
  Rev. Lett.}\ }\textbf {\bibinfo {volume} {85}},\ \bibinfo {pages} {503}
  (\bibinfo {year} {2000})}\BibitemShut {NoStop}%
\bibitem [{\citenamefont {Ninham}\ \emph {et~al.}(2014)\citenamefont {Ninham},
  \citenamefont {Bostr\"om}, \citenamefont {Persson}, \citenamefont {Brevik},
  \citenamefont {Buhmann},\ and\ \citenamefont {Sernelius}}]{EPJDNinham2014}%
  \BibitemOpen
  \bibfield  {author} {\bibinfo {author} {\bibfnamefont {B.~W.}\ \bibnamefont
  {Ninham}}, \bibinfo {author} {\bibfnamefont {M.}~\bibnamefont {Bostr\"om}},
  \bibinfo {author} {\bibfnamefont {C.}~\bibnamefont {Persson}}, \bibinfo
  {author} {\bibfnamefont {I.}~\bibnamefont {Brevik}}, \bibinfo {author}
  {\bibfnamefont {S.~Y.}\ \bibnamefont {Buhmann}}, \ and\ \bibinfo {author}
  {\bibfnamefont {Bo~E.}\ \bibnamefont {Sernelius}},\ }\bibfield  {title}
  {\enquote {\bibinfo {title} {Casimir forces in a plasma: Possible connections
  to yukawa potentials},}\ }\href {\doibase 10.1140/epjd/e2014-50484-8}
  {\bibfield  {journal} {\bibinfo  {journal} {Eur. Phys. J. D}\ }\textbf
  {\bibinfo {volume} {68}},\ \bibinfo {pages} {328} (\bibinfo {year}
  {2014})}\BibitemShut {NoStop}%
\bibitem [{\citenamefont {Dou}\ \emph {et~al.}(2014)\citenamefont {Dou},
  \citenamefont {Lou}, \citenamefont {Bostr\"om}, \citenamefont {Brevik},\ and\
  \citenamefont {Persson}}]{Maofeng2014PhysRevB.89.201407}%
  \BibitemOpen
  \bibfield  {author} {\bibinfo {author} {\bibfnamefont {Maofeng}\ \bibnamefont
  {Dou}}, \bibinfo {author} {\bibfnamefont {Fei}\ \bibnamefont {Lou}}, \bibinfo
  {author} {\bibfnamefont {Mathias}\ \bibnamefont {Bostr\"om}}, \bibinfo
  {author} {\bibfnamefont {Iver}\ \bibnamefont {Brevik}}, \ and\ \bibinfo
  {author} {\bibfnamefont {Clas}\ \bibnamefont {Persson}},\ }\bibfield  {title}
  {\enquote {\bibinfo {title} {Casimir quantum levitation tuned by means of
  material properties and geometries},}\ }\href {\doibase
  10.1103/PhysRevB.89.201407} {\bibfield  {journal} {\bibinfo  {journal} {Phys.
  Rev. B}\ }\textbf {\bibinfo {volume} {89}},\ \bibinfo {pages} {201407}
  (\bibinfo {year} {2014})}\BibitemShut {NoStop}%
\bibitem [{\citenamefont {Esteso}\ \emph {et~al.}(2015)\citenamefont {Esteso},
  \citenamefont {Carretero-Palacios},\ and\ \citenamefont
  {Miguez}}]{Estesodoi:10.1021/jp511851z}%
  \BibitemOpen
  \bibfield  {author} {\bibinfo {author} {\bibfnamefont {V.}~\bibnamefont
  {Esteso}}, \bibinfo {author} {\bibfnamefont {S.}~\bibnamefont
  {Carretero-Palacios}}, \ and\ \bibinfo {author} {\bibfnamefont
  {H.}~\bibnamefont {Miguez}},\ }\bibfield  {title} {\enquote {\bibinfo {title}
  {Nanolevitation phenomena in real plane-parallel systems due to the balance
  between casimir and gravity forces},}\ }\href {\doibase 10.1021/jp511851z}
  {\bibfield  {journal} {\bibinfo  {journal} {The Journal of Physical Chemistry
  C}\ }\textbf {\bibinfo {volume} {119}},\ \bibinfo {pages} {5663--5670}
  (\bibinfo {year} {2015})}\BibitemShut {NoStop}%
\bibitem [{\citenamefont {Bostr\"om}\ \emph {et~al.}(2018)\citenamefont
  {Bostr\"om}, \citenamefont {Dou}, \citenamefont {Malyi}, \citenamefont
  {Parashar}, \citenamefont {Parsons}, \citenamefont {Brevik},\ and\
  \citenamefont {Persson}}]{PhysRevB.97.125421}%
  \BibitemOpen
  \bibfield  {author} {\bibinfo {author} {\bibfnamefont {Mathias}\ \bibnamefont
  {Bostr\"om}}, \bibinfo {author} {\bibfnamefont {Maofeng}\ \bibnamefont
  {Dou}}, \bibinfo {author} {\bibfnamefont {Oleksandr~I.}\ \bibnamefont
  {Malyi}}, \bibinfo {author} {\bibfnamefont {Prachi}\ \bibnamefont
  {Parashar}}, \bibinfo {author} {\bibfnamefont {Drew~F.}\ \bibnamefont
  {Parsons}}, \bibinfo {author} {\bibfnamefont {Iver}\ \bibnamefont {Brevik}},
  \ and\ \bibinfo {author} {\bibfnamefont {Clas}\ \bibnamefont {Persson}},\
  }\bibfield  {title} {\enquote {\bibinfo {title} {{Fluid-sensitive nanoscale
  switching with quantum levitation controlled by
  $\ensuremath{\alpha}$-Sn/$\ensuremath{\beta}$-Sn phase transition}},}\ }\href
  {\doibase 10.1103/PhysRevB.97.125421} {\bibfield  {journal} {\bibinfo
  {journal} {Phys. Rev. B}\ }\textbf {\bibinfo {volume} {97}},\ \bibinfo
  {pages} {125421} (\bibinfo {year} {2018})}\BibitemShut {NoStop}%
\bibitem [{\citenamefont {Esteso}\ \emph {et~al.}(2019)\citenamefont {Esteso},
  \citenamefont {Carretero-Palacios},\ and\ \citenamefont
  {Miguez}}]{Estesosdoi:10.1021/acs.jpclett.9b02030}%
  \BibitemOpen
  \bibfield  {author} {\bibinfo {author} {\bibfnamefont {V.}~\bibnamefont
  {Esteso}}, \bibinfo {author} {\bibfnamefont {S.}~\bibnamefont
  {Carretero-Palacios}}, \ and\ \bibinfo {author} {\bibfnamefont
  {H.}~\bibnamefont {Miguez}},\ }\bibfield  {title} {\enquote {\bibinfo {title}
  {{C}asimir-{L}ifshitz force based optical resonators},}\ }\href {\doibase
  10.1021/acs.jpclett.9b02030} {\bibfield  {journal} {\bibinfo  {journal} {The
  Journal of Physical Chemistry Letters}\ }\textbf {\bibinfo {volume} {10}},\
  \bibinfo {pages} {5856--5860} (\bibinfo {year} {2019})}\BibitemShut {NoStop}%
\bibitem [{\citenamefont {Esteso}\ \emph {et~al.}(2020)\citenamefont {Esteso},
  \citenamefont {Carretero-Palacios}, \citenamefont {MacDowell}, \citenamefont
  {Fiedler}, \citenamefont {Parsons}, \citenamefont {Spallek}, \citenamefont
  {M\'iguez}, \citenamefont {Persson}, \citenamefont {Buhmann}, \citenamefont
  {Brevik},\ and\ \citenamefont {Bostr\"om}}]{Esteso4layerPCCP2020}%
  \BibitemOpen
  \bibfield  {author} {\bibinfo {author} {\bibfnamefont {V.}~\bibnamefont
  {Esteso}}, \bibinfo {author} {\bibfnamefont {S.}~\bibnamefont
  {Carretero-Palacios}}, \bibinfo {author} {\bibfnamefont {L.~G.}\ \bibnamefont
  {MacDowell}}, \bibinfo {author} {\bibfnamefont {J.}~\bibnamefont {Fiedler}},
  \bibinfo {author} {\bibfnamefont {D.~F.}\ \bibnamefont {Parsons}}, \bibinfo
  {author} {\bibfnamefont {F.}~\bibnamefont {Spallek}}, \bibinfo {author}
  {\bibfnamefont {H.}~\bibnamefont {M\'iguez}}, \bibinfo {author}
  {\bibfnamefont {C.}~\bibnamefont {Persson}}, \bibinfo {author} {\bibfnamefont
  {S.~Y.}\ \bibnamefont {Buhmann}}, \bibinfo {author} {\bibfnamefont
  {I.}~\bibnamefont {Brevik}}, \ and\ \bibinfo {author} {\bibfnamefont
  {M.}~\bibnamefont {Bostr\"om}},\ }\bibfield  {title} {\enquote {\bibinfo
  {title} {Premelting of ice adsorbed on a rock surface},}\ }\href {\doibase
  10.1039/C9CP06836H} {\bibfield  {journal} {\bibinfo  {journal} {Phys. Chem.
  Chem. Phys.}\ }\textbf {\bibinfo {volume} {22}},\ \bibinfo {pages}
  {11362--11373} (\bibinfo {year} {2020})}\BibitemShut {NoStop}%
\bibitem [{\citenamefont {Ninham}\ \emph {et~al.}(2022)\citenamefont {Ninham},
  \citenamefont {Brevik},\ and\ \citenamefont
  {Bostr\"om}}]{Ninham_Brevik_Bostrom_2022}%
  \BibitemOpen
  \bibfield  {author} {\bibinfo {author} {\bibfnamefont {B.~W.}\ \bibnamefont
  {Ninham}}, \bibinfo {author} {\bibfnamefont {I.}~\bibnamefont {Brevik}}, \
  and\ \bibinfo {author} {\bibfnamefont {M.}~\bibnamefont {Bostr\"om}},\
  }\bibfield  {title} {\enquote {\bibinfo {title} {Equivalence of
  electromagnetic fluctuation and nuclear (yukawa) forces: the $\pi_0$ meson,
  its mass and lifetime},}\ }\href {\doibase 10.36253/Substantia-1807}
  {\bibfield  {journal} {\bibinfo  {journal} {Substantia}\ }\textbf {\bibinfo
  {volume} {7}},\ \bibinfo {pages} {7--14} (\bibinfo {year}
  {2022})}\BibitemShut {NoStop}%
\bibitem [{\citenamefont {Li}\ \emph {et~al.}(2023)\citenamefont {Li},
  \citenamefont {Brevik}, \citenamefont {Malyi},\ and\ \citenamefont
  {Bostr\"om}}]{LiBrevikMalyiBostromPhysRevE.108.034801}%
  \BibitemOpen
  \bibfield  {author} {\bibinfo {author} {\bibfnamefont {Y.}~\bibnamefont
  {Li}}, \bibinfo {author} {\bibfnamefont {I.}~\bibnamefont {Brevik}}, \bibinfo
  {author} {\bibfnamefont {O.~I.}\ \bibnamefont {Malyi}}, \ and\ \bibinfo
  {author} {\bibfnamefont {M.}~\bibnamefont {Bostr\"om}},\ }\bibfield  {title}
  {\enquote {\bibinfo {title} {Different pathways to anomalous stabilization of
  ice layers on methane hydrates},}\ }\href {\doibase
  10.1103/PhysRevE.108.034801} {\bibfield  {journal} {\bibinfo  {journal}
  {Phys. Rev. E}\ }\textbf {\bibinfo {volume} {108}},\ \bibinfo {pages}
  {034801} (\bibinfo {year} {2023})}\BibitemShut {NoStop}%
\bibitem [{\citenamefont {Bostr\"om}\ \emph
  {et~al.}(2023{\natexlab{a}})\citenamefont {Bostr\"om}, \citenamefont {Khan},
  \citenamefont {Gopidi}, \citenamefont {Brevik}, \citenamefont {Li},
  \citenamefont {Persson},\ and\ \citenamefont
  {Malyi}}]{BostromRizwanHarshanBrevikLiPerssonMalyi2023spontaneous}%
  \BibitemOpen
  \bibfield  {author} {\bibinfo {author} {\bibfnamefont {M.}~\bibnamefont
  {Bostr\"om}}, \bibinfo {author} {\bibfnamefont {M.~R.}\ \bibnamefont {Khan}},
  \bibinfo {author} {\bibfnamefont {H.~R.}\ \bibnamefont {Gopidi}}, \bibinfo
  {author} {\bibfnamefont {I.}~\bibnamefont {Brevik}}, \bibinfo {author}
  {\bibfnamefont {Y.}~\bibnamefont {Li}}, \bibinfo {author} {\bibfnamefont
  {C.}~\bibnamefont {Persson}}, \ and\ \bibinfo {author} {\bibfnamefont
  {O.~I.}\ \bibnamefont {Malyi}},\ }\bibfield  {title} {\enquote {\bibinfo
  {title} {Tuning the {Casimir-L}ifshitz force with gapped metals},}\ }\href
  {\doibase 10.1103/PhysRevB.108.165306} {\bibfield  {journal} {\bibinfo
  {journal} {Phys. Rev. B}\ }\textbf {\bibinfo {volume} {108}},\ \bibinfo
  {pages} {165306} (\bibinfo {year} {2023}{\natexlab{a}})}\BibitemShut
  {NoStop}%
\bibitem [{\citenamefont {Klimchitskaya}\ and\ \citenamefont
  {Mostepanenko}(2023)}]{MostKlim2023}%
  \BibitemOpen
  \bibfield  {author} {\bibinfo {author} {\bibfnamefont {Galina~L.}\
  \bibnamefont {Klimchitskaya}}\ and\ \bibinfo {author} {\bibfnamefont
  {Vladimir~M.}\ \bibnamefont {Mostepanenko}},\ }\bibfield  {title} {\enquote
  {\bibinfo {title} {Casimir effect invalidates the drude model for transverse
  electric evanescent waves},}\ }\href {\doibase 10.3390/physics5040062}
  {\bibfield  {journal} {\bibinfo  {journal} {Physics}\ }\textbf {\bibinfo
  {volume} {5}},\ \bibinfo {pages} {952--967} (\bibinfo {year}
  {2023})}\BibitemShut {NoStop}%
\bibitem [{\citenamefont {Hauxwell}\ and\ \citenamefont
  {Ottewill}(1970)}]{Haux}%
  \BibitemOpen
  \bibfield  {author} {\bibinfo {author} {\bibfnamefont {F}~\bibnamefont
  {Hauxwell}}\ and\ \bibinfo {author} {\bibfnamefont {R.H}\ \bibnamefont
  {Ottewill}},\ }\bibfield  {title} {\enquote {\bibinfo {title} {{A study of
  the surface of water by hydrocarbon adsorption}},}\ }\href {\doibase
  https://doi.org/10.1016/0021-9797(70)90208-0} {\bibfield  {journal} {\bibinfo
   {journal} {Journal of Colloid and Interface Science}\ }\textbf {\bibinfo
  {volume} {34}},\ \bibinfo {pages} {473 -- 479} (\bibinfo {year}
  {1970})}\BibitemShut {NoStop}%
\bibitem [{\citenamefont {Anderson}\ and\ \citenamefont
  {Sabisky}(1970)}]{AndSab}%
  \BibitemOpen
  \bibfield  {author} {\bibinfo {author} {\bibfnamefont {C.~H.}\ \bibnamefont
  {Anderson}}\ and\ \bibinfo {author} {\bibfnamefont {E.~S.}\ \bibnamefont
  {Sabisky}},\ }\bibfield  {title} {\enquote {\bibinfo {title} {{Phonon
  Interference in Thin Films of Liquid Helium}},}\ }\href {\doibase
  10.1103/PhysRevLett.24.1049} {\bibfield  {journal} {\bibinfo  {journal}
  {Phys. Rev. Lett.}\ }\textbf {\bibinfo {volume} {24}},\ \bibinfo {pages}
  {1049--1052} (\bibinfo {year} {1970})}\BibitemShut {NoStop}%
\bibitem [{\citenamefont {Lamoreaux}(1997)}]{Lamo1997}%
  \BibitemOpen
  \bibfield  {author} {\bibinfo {author} {\bibfnamefont {S.~K.}\ \bibnamefont
  {Lamoreaux}},\ }\bibfield  {title} {\enquote {\bibinfo {title}
  {{Demonstration of the Casimir Force in the 0.6 to 6\,$\mu$\,m Range}},}\
  }\href {\doibase 10.1103/PhysRevLett.78.5} {\bibfield  {journal} {\bibinfo
  {journal} {Phys. Rev. Lett.}\ }\textbf {\bibinfo {volume} {78}},\ \bibinfo
  {pages} {5} (\bibinfo {year} {1997})}\BibitemShut {NoStop}%
\bibitem [{\citenamefont {Harris}\ \emph {et~al.}(2000)\citenamefont {Harris},
  \citenamefont {Chen},\ and\ \citenamefont
  {Mohideen}}]{HarrisPhysRevA.62.052109}%
  \BibitemOpen
  \bibfield  {author} {\bibinfo {author} {\bibfnamefont {B.~W.}\ \bibnamefont
  {Harris}}, \bibinfo {author} {\bibfnamefont {F.}~\bibnamefont {Chen}}, \ and\
  \bibinfo {author} {\bibfnamefont {U.}~\bibnamefont {Mohideen}},\ }\bibfield
  {title} {\enquote {\bibinfo {title} {Precision measurement of the {C}asimir
  force using gold surfaces},}\ }\href {\doibase 10.1103/PhysRevA.62.052109}
  {\bibfield  {journal} {\bibinfo  {journal} {Phys. Rev. A}\ }\textbf {\bibinfo
  {volume} {62}},\ \bibinfo {pages} {052109} (\bibinfo {year}
  {2000})}\BibitemShut {NoStop}%
\bibitem [{\citenamefont {Decca}\ \emph {et~al.}(2003)\citenamefont {Decca},
  \citenamefont {L\'opez}, \citenamefont {Fischbach},\ and\ \citenamefont
  {Krause}}]{DeccaPhysRevLett.91.050402}%
  \BibitemOpen
  \bibfield  {author} {\bibinfo {author} {\bibfnamefont {R.~S.}\ \bibnamefont
  {Decca}}, \bibinfo {author} {\bibfnamefont {D.}~\bibnamefont {L\'opez}},
  \bibinfo {author} {\bibfnamefont {E.}~\bibnamefont {Fischbach}}, \ and\
  \bibinfo {author} {\bibfnamefont {D.~E.}\ \bibnamefont {Krause}},\ }\bibfield
   {title} {\enquote {\bibinfo {title} {Measurement of the {C}asimir force
  between dissimilar metals},}\ }\href {\doibase 10.1103/PhysRevLett.91.050402}
  {\bibfield  {journal} {\bibinfo  {journal} {Phys. Rev. Lett.}\ }\textbf
  {\bibinfo {volume} {91}},\ \bibinfo {pages} {050402} (\bibinfo {year}
  {2003})}\BibitemShut {NoStop}%
\bibitem [{\citenamefont {Feiler}\ \emph {et~al.}(2008)\citenamefont {Feiler},
  \citenamefont {Bergstr{\"o}m},\ and\ \citenamefont {Rutland}}]{Feiler2008}%
  \BibitemOpen
  \bibfield  {author} {\bibinfo {author} {\bibfnamefont {A.~A.}\ \bibnamefont
  {Feiler}}, \bibinfo {author} {\bibfnamefont {L.}~\bibnamefont
  {Bergstr{\"o}m}}, \ and\ \bibinfo {author} {\bibfnamefont {M.~W.}\
  \bibnamefont {Rutland}},\ }\bibfield  {title} {\enquote {\bibinfo {title}
  {{Superlubricity Using Repulsive van der Waals Forces}},}\ }\href {\doibase
  https://doi.org/10.1021/la7036907} {\bibfield  {journal} {\bibinfo  {journal}
  {Langmuir}\ }\textbf {\bibinfo {volume} {24}},\ \bibinfo {pages} {2274 --
  2276} (\bibinfo {year} {2008})}\BibitemShut {NoStop}%
\bibitem [{Mun()}]{Munday2009}%
  \BibitemOpen
  \href@noop {} {\ }\BibitemShut {NoStop}%
\bibitem [{\citenamefont {Somers}\ \emph {et~al.}(2018)\citenamefont {Somers},
  \citenamefont {Garrett}, \citenamefont {Palm},\ and\ \citenamefont
  {Munday}}]{SomersGarrettPalmMunday_CasimirTorque}%
  \BibitemOpen
  \bibfield  {author} {\bibinfo {author} {\bibfnamefont {D.A.T.}\ \bibnamefont
  {Somers}}, \bibinfo {author} {\bibfnamefont {J.L.}\ \bibnamefont {Garrett}},
  \bibinfo {author} {\bibfnamefont {K.J.}\ \bibnamefont {Palm}}, \ and\
  \bibinfo {author} {\bibfnamefont {J.~N.}\ \bibnamefont {Munday}},\ }\bibfield
   {title} {\enquote {\bibinfo {title} {Measurement of the {Casimir} torque},}\
  }\href {\doibase 10.1038/s41586-018-0777-8} {\bibfield  {journal} {\bibinfo
  {journal} {Nature}\ }\textbf {\bibinfo {volume} {564}},\ \bibinfo {pages}
  {386--389} (\bibinfo {year} {2018})}\BibitemShut {NoStop}%
\bibitem [{\citenamefont {Mahanty}\ and\ \citenamefont {Ninham}(1976)}]{Maha}%
  \BibitemOpen
  \bibfield  {author} {\bibinfo {author} {\bibfnamefont {J.}~\bibnamefont
  {Mahanty}}\ and\ \bibinfo {author} {\bibfnamefont {B.~W.}\ \bibnamefont
  {Ninham}},\ }\href {\href{https://doi.org/10.1002/bbpc.19770810816}} {\emph
  {\bibinfo {title} {{Dispersion Forces}}}}\ (\bibinfo  {publisher} {Academic
  Press},\ \bibinfo {address} {London},\ \bibinfo {year} {1976})\BibitemShut
  {NoStop}%
\bibitem [{\citenamefont {Milton}(2001)}]{milton01}%
  \BibitemOpen
  \bibfield  {author} {\bibinfo {author} {\bibfnamefont {Kimball~A}\
  \bibnamefont {Milton}},\ }\href@noop {} {\emph {\bibinfo {title} {The Casimir
  effect: physical manifestations of zero-point energy}}}\ (\bibinfo
  {publisher} {World Scientific},\ \bibinfo {year} {2001})\BibitemShut
  {NoStop}%
\bibitem [{\citenamefont {Parsegian}(2006)}]{Pars}%
  \BibitemOpen
  \bibfield  {author} {\bibinfo {author} {\bibfnamefont {V.~A.}\ \bibnamefont
  {Parsegian}},\ }\href@noop {} {\emph {\bibinfo {title} {Van der Waals forces:
  A handbook for biologists, chemists, engineers, and physicists}}}\ (\bibinfo
  {publisher} {Cambridge University Press},\ \bibinfo {address} {New York},\
  \bibinfo {year} {2006})\BibitemShut {NoStop}%
\bibitem [{\citenamefont {Bordag}\ \emph {et~al.}(2009)\citenamefont {Bordag},
  \citenamefont {Klimchitskaya}, \citenamefont {Mohideen},\ and\ \citenamefont
  {Mostepanenko}}]{Bordagbook}%
  \BibitemOpen
  \bibfield  {author} {\bibinfo {author} {\bibfnamefont {M.}~\bibnamefont
  {Bordag}}, \bibinfo {author} {\bibfnamefont {G.~L.}\ \bibnamefont
  {Klimchitskaya}}, \bibinfo {author} {\bibfnamefont {U.}~\bibnamefont
  {Mohideen}}, \ and\ \bibinfo {author} {\bibfnamefont {V.~M.}\ \bibnamefont
  {Mostepanenko}},\ }\href@noop {} {\emph {\bibinfo {title} {{Advances in the
  Casimir Effect}}}}\ (\bibinfo  {publisher} {Oxford Science Publications},\
  \bibinfo {address} {Oxford},\ \bibinfo {year} {2009})\BibitemShut {NoStop}%
\bibitem [{\citenamefont {Sernelius}(2005)}]{Ser}%
  \BibitemOpen
  \bibfield  {author} {\bibinfo {author} {\bibfnamefont {Bo~E.}\ \bibnamefont
  {Sernelius}},\ }\href {\doibase 10.1002/3527603166} {\emph {\bibinfo {title}
  {{Surface Modes in Physics}}}}\ (\bibinfo  {publisher} {John Wiley \& Sons,
  Inc.},\ \bibinfo {year} {2005})\BibitemShut {NoStop}%
\bibitem [{\citenamefont {Buhmann}(2012)}]{Buhmann12a}%
  \BibitemOpen
  \bibfield  {author} {\bibinfo {author} {\bibfnamefont {S.~Y.}\ \bibnamefont
  {Buhmann}},\ }\href@noop {} {\emph {\bibinfo {title} {{Dispersion Forces I:
  Macroscopic quantum electrodynamics and ground-state Casimir, Casimir--Polder
  and van der Waals forces}}}}\ (\bibinfo  {publisher} {Springer},\ \bibinfo
  {address} {Heidelberg},\ \bibinfo {year} {2012})\BibitemShut {NoStop}%
\bibitem [{\citenamefont {Sernelius}(2018)}]{Ser2018}%
  \BibitemOpen
  \bibfield  {author} {\bibinfo {author} {\bibfnamefont {Bo~E.}\ \bibnamefont
  {Sernelius}},\ }\href {\doibase 10.1007/978-3-319-99831-2} {\emph {\bibinfo
  {title} {{Fundamentals of van der Waals and Casimir Interactions}}}},\
  Springer Series on Atomic, Optical, and Plasma Physics\ (\bibinfo
  {publisher} {Springer International Publishing},\ \bibinfo {year}
  {2018})\BibitemShut {NoStop}%
\bibitem [{\citenamefont {Bostr\"om}\ \emph {et~al.}(2001)\citenamefont
  {Bostr\"om}, \citenamefont {Williams},\ and\ \citenamefont
  {Ninham}}]{Bost2001}%
  \BibitemOpen
  \bibfield  {author} {\bibinfo {author} {\bibfnamefont {M.}~\bibnamefont
  {Bostr\"om}}, \bibinfo {author} {\bibfnamefont {D.~R.~M.}\ \bibnamefont
  {Williams}}, \ and\ \bibinfo {author} {\bibfnamefont {B.~W.}\ \bibnamefont
  {Ninham}},\ }\bibfield  {title} {\enquote {\bibinfo {title} {{Specific Ion
  Effects: Why DLVO Theory Fails for Biology and Colloid Systems}},}\ }\href
  {\doibase 10.1103/PhysRevLett.87.168103} {\bibfield  {journal} {\bibinfo
  {journal} {Phys. Rev. Lett.}\ }\textbf {\bibinfo {volume} {87}},\ \bibinfo
  {pages} {168103} (\bibinfo {year} {2001})}\BibitemShut {NoStop}%
\bibitem [{\citenamefont {Parsons}\ and\ \citenamefont
  {Ninham}(2009)}]{ParsonsNinham2009}%
  \BibitemOpen
  \bibfield  {author} {\bibinfo {author} {\bibfnamefont {Drew~F.}\ \bibnamefont
  {Parsons}}\ and\ \bibinfo {author} {\bibfnamefont {Barry~W.}\ \bibnamefont
  {Ninham}},\ }\bibfield  {title} {\enquote {\bibinfo {title} {{Ab Initio Molar
  Volumes and Gaussian Radii}},}\ }\href {\doibase 10.1021/jp802984b}
  {\bibfield  {journal} {\bibinfo  {journal} {J. Phys. Chem. A}\ }\textbf
  {\bibinfo {volume} {113}},\ \bibinfo {pages} {1141--1150} (\bibinfo {year}
  {2009})},\ \Eprint
  {http://arxiv.org/abs/http://pubs.acs.org/doi/pdf/10.1021/jp802984b}
  {http://pubs.acs.org/doi/pdf/10.1021/jp802984b} \BibitemShut {NoStop}%
\bibitem [{\citenamefont {Parsons}\ and\ \citenamefont
  {Ninham}(2012)}]{ParsonsNinham2012}%
  \BibitemOpen
  \bibfield  {author} {\bibinfo {author} {\bibfnamefont {D.~F.}\ \bibnamefont
  {Parsons}}\ and\ \bibinfo {author} {\bibfnamefont {B.~W.}\ \bibnamefont
  {Ninham}},\ }\bibfield  {title} {\enquote {\bibinfo {title} {Nonelectrostatic
  ionic forces between dissimilar surfaces: A mechanism for colloid
  separation},}\ }\href {\doibase 10.1021/jp212154c} {\bibfield  {journal}
  {\bibinfo  {journal} {J. Phys. Chem. C}\ }\textbf {\bibinfo {volume} {116}},\
  \bibinfo {pages} {7782--7792} (\bibinfo {year} {2012})}\BibitemShut {NoStop}%
\bibitem [{\citenamefont {Medda}\ \emph {et~al.}(2012)\citenamefont {Medda},
  \citenamefont {Barse}, \citenamefont {Cugia}, \citenamefont {Bostr\"om},
  \citenamefont {Parsons}, \citenamefont {Ninham}, \citenamefont {Monduzzi},\
  and\ \citenamefont {Salis}}]{Medda2012doi:10.1021/la3035984}%
  \BibitemOpen
  \bibfield  {author} {\bibinfo {author} {\bibfnamefont {L.}~\bibnamefont
  {Medda}}, \bibinfo {author} {\bibfnamefont {B.}~\bibnamefont {Barse}},
  \bibinfo {author} {\bibfnamefont {F.}~\bibnamefont {Cugia}}, \bibinfo
  {author} {\bibfnamefont {M.}~\bibnamefont {Bostr\"om}}, \bibinfo {author}
  {\bibfnamefont {D.~F.}\ \bibnamefont {Parsons}}, \bibinfo {author}
  {\bibfnamefont {B.~W.}\ \bibnamefont {Ninham}}, \bibinfo {author}
  {\bibfnamefont {M.}~\bibnamefont {Monduzzi}}, \ and\ \bibinfo {author}
  {\bibfnamefont {A.}~\bibnamefont {Salis}},\ }\bibfield  {title} {\enquote
  {\bibinfo {title} {Hofmeister challenges: Ion binding and charge of the bsa
  protein as explicit examples},}\ }\href {\doibase 10.1021/la3035984}
  {\bibfield  {journal} {\bibinfo  {journal} {Langmuir}\ }\textbf {\bibinfo
  {volume} {28}},\ \bibinfo {pages} {16355--16363} (\bibinfo {year}
  {2012})}\BibitemShut {NoStop}%
\bibitem [{\citenamefont {Duignan}\ \emph {et~al.}(2013)\citenamefont
  {Duignan}, \citenamefont {Parsons},\ and\ \citenamefont
  {Ninham}}]{DuignanNinhamParsons2013solvation}%
  \BibitemOpen
  \bibfield  {author} {\bibinfo {author} {\bibfnamefont {Timothy~T.}\
  \bibnamefont {Duignan}}, \bibinfo {author} {\bibfnamefont {Drew~F.}\
  \bibnamefont {Parsons}}, \ and\ \bibinfo {author} {\bibfnamefont {Barry~W.}\
  \bibnamefont {Ninham}},\ }\bibfield  {title} {\enquote {\bibinfo {title} {{A
  Continuum Model of Solvation Energies Including Electrostatic, Dispersion,
  and Cavity Contributions}},}\ }\href {\doibase 10.1021/jp403596c} {\bibfield
  {journal} {\bibinfo  {journal} {J. Phys. Chem. B}\ }\textbf {\bibinfo
  {volume} {117}},\ \bibinfo {pages} {9421--9429} (\bibinfo {year} {2013})},\
  \Eprint {http://arxiv.org/abs/http://pubs.acs.org/doi/pdf/10.1021/jp403596c}
  {http://pubs.acs.org/doi/pdf/10.1021/jp403596c} \BibitemShut {NoStop}%
\bibitem [{\citenamefont {Parsons}\ and\ \citenamefont
  {Salis}(2015)}]{ParsonsSalis2015}%
  \BibitemOpen
  \bibfield  {author} {\bibinfo {author} {\bibfnamefont {D.~F.}\ \bibnamefont
  {Parsons}}\ and\ \bibinfo {author} {\bibfnamefont {A.}~\bibnamefont
  {Salis}},\ }\bibfield  {title} {\enquote {\bibinfo {title} {The impact of the
  competitive adsorption of ions at surface sites on surface free energies and
  surface forces},}\ }\href {\doibase 10.1063/1.4916519} {\bibfield  {journal}
  {\bibinfo  {journal} {J. Chem. Phys.}\ }\textbf {\bibinfo {volume} {142}},\
  \bibinfo {pages} {134707} (\bibinfo {year} {2015})}\BibitemShut {NoStop}%
\bibitem [{\citenamefont {Bostr{\"o}m}\ \emph {et~al.}(2016)\citenamefont
  {Bostr{\"o}m}, \citenamefont {Malyi}, \citenamefont {Thiyam}, \citenamefont
  {Berland}, \citenamefont {Brevik}, \citenamefont {Persson},\ and\
  \citenamefont {Parsons}}]{Bostr2016}%
  \BibitemOpen
  \bibfield  {author} {\bibinfo {author} {\bibfnamefont {M.}~\bibnamefont
  {Bostr{\"o}m}}, \bibinfo {author} {\bibfnamefont {O.~I.}\ \bibnamefont
  {Malyi}}, \bibinfo {author} {\bibfnamefont {P.}~\bibnamefont {Thiyam}},
  \bibinfo {author} {\bibfnamefont {K.}~\bibnamefont {Berland}}, \bibinfo
  {author} {\bibfnamefont {I.}~\bibnamefont {Brevik}}, \bibinfo {author}
  {\bibfnamefont {C.}~\bibnamefont {Persson}}, \ and\ \bibinfo {author}
  {\bibfnamefont {D.~F.}\ \bibnamefont {Parsons}},\ }\bibfield  {title}
  {\enquote {\bibinfo {title} {The influence of lifshitz forces and gas on
  premelting of ice within porous materials},}\ }\href
  {http://stacks.iop.org/0295-5075/115/i=1/a=13001} {\bibfield  {journal}
  {\bibinfo  {journal} {EPL (Europhys. Lett.)}\ }\textbf {\bibinfo {volume}
  {115}},\ \bibinfo {pages} {13001} (\bibinfo {year} {2016})}\BibitemShut
  {NoStop}%
\bibitem [{\citenamefont {Thiyam}\ \emph {et~al.}(2018)\citenamefont {Thiyam},
  \citenamefont {Fiedler}, \citenamefont {Buhmann}, \citenamefont {Persson},
  \citenamefont {Brevik}, \citenamefont {Bostr{\"o}m},\ and\ \citenamefont
  {Parsons}}]{ThiyamFiedlerBuhmannPerssonBrevikBostromParsons2018}%
  \BibitemOpen
  \bibfield  {author} {\bibinfo {author} {\bibfnamefont {P.}~\bibnamefont
  {Thiyam}}, \bibinfo {author} {\bibfnamefont {J.}~\bibnamefont {Fiedler}},
  \bibinfo {author} {\bibfnamefont {S.~Y.}\ \bibnamefont {Buhmann}}, \bibinfo
  {author} {\bibfnamefont {C.}~\bibnamefont {Persson}}, \bibinfo {author}
  {\bibfnamefont {I.}~\bibnamefont {Brevik}}, \bibinfo {author} {\bibfnamefont
  {M.}~\bibnamefont {Bostr{\"o}m}}, \ and\ \bibinfo {author} {\bibfnamefont
  {D.~F.}\ \bibnamefont {Parsons}},\ }\bibfield  {title} {\enquote {\bibinfo
  {title} {{Ice Particles Sink below the Water Surface Due to a Balance of
  Salt, van der Waals, and Buoyancy Forces}},}\ }\href {\doibase
  10.1021/acs.jpcc.8b02351} {\bibfield  {journal} {\bibinfo  {journal} {J.
  Phys. Chem. C}\ }\textbf {\bibinfo {volume} {122}},\ \bibinfo {pages}
  {15311--15317} (\bibinfo {year} {2018})}\BibitemShut {NoStop}%
\bibitem [{\citenamefont {Ninham}\ and\ \citenamefont
  {Yaminsky}(1997)}]{NinhamYaminsky1997}%
  \BibitemOpen
  \bibfield  {author} {\bibinfo {author} {\bibfnamefont {Barry~W.}\
  \bibnamefont {Ninham}}\ and\ \bibinfo {author} {\bibfnamefont {Vassili}\
  \bibnamefont {Yaminsky}},\ }\bibfield  {title} {\enquote {\bibinfo {title}
  {Ion binding and ion specificity: The hofmeister effect and onsager and
  lifshitz theories},}\ }\href {\doibase 10.1021/la960974y} {\bibfield
  {journal} {\bibinfo  {journal} {Langmuir}\ }\textbf {\bibinfo {volume}
  {13}},\ \bibinfo {pages} {2097--2108} (\bibinfo {year} {1997})},\ \Eprint
  {http://arxiv.org/abs/https://doi.org/10.1021/la960974y}
  {https://doi.org/10.1021/la960974y} \BibitemShut {NoStop}%
\bibitem [{\citenamefont {Bostr\"om}\ \emph
  {et~al.}(2023{\natexlab{b}})\citenamefont {Bostr\"om}, \citenamefont {Li},
  \citenamefont {Brevik}, \citenamefont {Persson}, \citenamefont
  {Carretero-Palacios},\ and\ \citenamefont
  {Malyi}}]{BostromvdWicegrowthmistPCCP2023}%
  \BibitemOpen
  \bibfield  {author} {\bibinfo {author} {\bibfnamefont {M.}~\bibnamefont
  {Bostr\"om}}, \bibinfo {author} {\bibfnamefont {Y.}~\bibnamefont {Li}},
  \bibinfo {author} {\bibfnamefont {I.}~\bibnamefont {Brevik}}, \bibinfo
  {author} {\bibfnamefont {C.}~\bibnamefont {Persson}}, \bibinfo {author}
  {\bibfnamefont {S.}~\bibnamefont {Carretero-Palacios}}, \ and\ \bibinfo
  {author} {\bibfnamefont {O.~I.}\ \bibnamefont {Malyi}},\ }\bibfield  {title}
  {\enquote {\bibinfo {title} {van der waals induced ice growth on partially
  melted ice nuclei in mist and fog},}\ }\href {\doibase 10.1039/D3CP04157C}
  {\bibfield  {journal} {\bibinfo  {journal} {Phys. Chem. Chem. Phys.}\
  }\textbf {\bibinfo {volume} {25}},\ \bibinfo {pages} {32709--32714} (\bibinfo
  {year} {2023}{\natexlab{b}})}\BibitemShut {NoStop}%
\bibitem [{\citenamefont {Luengo-M\'arquez}\ and\ \citenamefont
  {MacDowell}(2021)}]{LUENGOMARQUEZMacDowell2021}%
  \BibitemOpen
  \bibfield  {author} {\bibinfo {author} {\bibfnamefont {J.}~\bibnamefont
  {Luengo-M\'arquez}}\ and\ \bibinfo {author} {\bibfnamefont {L.~G.}\
  \bibnamefont {MacDowell}},\ }\bibfield  {title} {\enquote {\bibinfo {title}
  {Lifshitz theory of wetting films at three phase coexistence: The case of ice
  nucleation on silver iodide (agi)},}\ }\href {\doibase
  https://doi.org/10.1016/j.jcis.2021.01.060} {\bibfield  {journal} {\bibinfo
  {journal} {J. Coll. Interf. Sci.}\ }\textbf {\bibinfo {volume} {590}},\
  \bibinfo {pages} {527--538} (\bibinfo {year} {2021})}\BibitemShut {NoStop}%
\bibitem [{\citenamefont {Luengo-Marquez}\ \emph {et~al.}(2022)\citenamefont
  {Luengo-Marquez}, \citenamefont {Izquierdo-Ruiz},\ and\ \citenamefont
  {MacDowell}}]{LuengoMarquez_IzquierdoRuiz_MacDowell2022}%
  \BibitemOpen
  \bibfield  {author} {\bibinfo {author} {\bibfnamefont {J.}~\bibnamefont
  {Luengo-Marquez}}, \bibinfo {author} {\bibfnamefont {F.}~\bibnamefont
  {Izquierdo-Ruiz}}, \ and\ \bibinfo {author} {\bibfnamefont {L.~G.}\
  \bibnamefont {MacDowell}},\ }\bibfield  {title} {\enquote {\bibinfo {title}
  {Intermolecular forces at ice and water interfaces: Premelting, surface
  freezing, and regelation},}\ }\href {\doibase 10.1063/5.0097378} {\bibfield
  {journal} {\bibinfo  {journal} {J. Chem. Phys.}\ }\textbf {\bibinfo {volume}
  {157}},\ \bibinfo {pages} {044704} (\bibinfo {year} {2022})}\BibitemShut
  {NoStop}%
\bibitem [{\citenamefont {Sikivie}(1983)}]{sikivie83}%
  \BibitemOpen
  \bibfield  {author} {\bibinfo {author} {\bibfnamefont {Pierre}\ \bibnamefont
  {Sikivie}},\ }\bibfield  {title} {\enquote {\bibinfo {title} {Experimental
  tests of the "invisible" axion},}\ }\href {\doibase
  10.1103/PhysRevLett.51.1415} {\bibfield  {journal} {\bibinfo  {journal}
  {Phys. Rev. Lett.}\ }\textbf {\bibinfo {volume} {51}},\ \bibinfo {pages}
  {1415} (\bibinfo {year} {1983})}\BibitemShut {NoStop}%
\bibitem [{\citenamefont {Weinberg}(1978{\natexlab{b}})}]{weinberg78}%
  \BibitemOpen
  \bibfield  {author} {\bibinfo {author} {\bibfnamefont {Steven}\ \bibnamefont
  {Weinberg}},\ }\bibfield  {title} {\enquote {\bibinfo {title} {A new light
  boson?}}\ }\href {\doibase 10.1103/PhysRevLett.40.223} {\bibfield  {journal}
  {\bibinfo  {journal} {Phys. Rev. Lett.}\ }\textbf {\bibinfo {volume} {40}},\
  \bibinfo {pages} {223} (\bibinfo {year} {1978}{\natexlab{b}})}\BibitemShut
  {NoStop}%
\bibitem [{\citenamefont {Preskill}\ \emph {et~al.}(1983)\citenamefont
  {Preskill}, \citenamefont {Wise},\ and\ \citenamefont
  {Wilczek}}]{preskill83}%
  \BibitemOpen
  \bibfield  {author} {\bibinfo {author} {\bibfnamefont {John}\ \bibnamefont
  {Preskill}}, \bibinfo {author} {\bibfnamefont {Mark~B}\ \bibnamefont {Wise}},
  \ and\ \bibinfo {author} {\bibfnamefont {Frank}\ \bibnamefont {Wilczek}},\
  }\bibfield  {title} {\enquote {\bibinfo {title} {Cosmology of the invisible
  axion},}\ }\href {\doibase 10.1016/0370-2693(83)90637-8} {\bibfield
  {journal} {\bibinfo  {journal} {Phys. Lett. B}\ }\textbf {\bibinfo {volume}
  {120}},\ \bibinfo {pages} {127--132} (\bibinfo {year} {1983})}\BibitemShut
  {NoStop}%
\bibitem [{\citenamefont {Abbott}\ and\ \citenamefont
  {Sikivie}(1983)}]{abbott83}%
  \BibitemOpen
  \bibfield  {author} {\bibinfo {author} {\bibfnamefont {Laurence~F}\
  \bibnamefont {Abbott}}\ and\ \bibinfo {author} {\bibfnamefont
  {P}~\bibnamefont {Sikivie}},\ }\bibfield  {title} {\enquote {\bibinfo {title}
  {A cosmological bound on the invisible axion},}\ }\href {\doibase
  10.1016/0370-2693(83)90638-X} {\bibfield  {journal} {\bibinfo  {journal}
  {Phys. Lett. B}\ }\textbf {\bibinfo {volume} {120}},\ \bibinfo {pages}
  {133--136} (\bibinfo {year} {1983})}\BibitemShut {NoStop}%
\bibitem [{\citenamefont {Dine}\ and\ \citenamefont {Fischler}(1983)}]{dine83}%
  \BibitemOpen
  \bibfield  {author} {\bibinfo {author} {\bibfnamefont {Michael}\ \bibnamefont
  {Dine}}\ and\ \bibinfo {author} {\bibfnamefont {Willy}\ \bibnamefont
  {Fischler}},\ }\bibfield  {title} {\enquote {\bibinfo {title} {The
  not-so-harmless axion},}\ }\href {\doibase 10.1016/0370-2693(83)90639-1}
  {\bibfield  {journal} {\bibinfo  {journal} {Phys. Lett. B}\ }\textbf
  {\bibinfo {volume} {120}},\ \bibinfo {pages} {137--141} (\bibinfo {year}
  {1983})}\BibitemShut {NoStop}%
\bibitem [{\citenamefont {Sikivie}(2008)}]{sikivie08}%
  \BibitemOpen
  \bibfield  {author} {\bibinfo {author} {\bibfnamefont {P.}~\bibnamefont
  {Sikivie}},\ }\href {\doibase 10.48550/arXiv.astro-ph/0610440} {\emph
  {\bibinfo {title} {Axion cosmology (Editors M. Kuster and G. Raffelt and B.
  Beltran)}}},\ edited by\ \bibinfo {editor} {\bibfnamefont {M.}~\bibnamefont
  {Kuster}}, \bibinfo {editor} {\bibfnamefont {G.}~\bibnamefont {Raffelt}}, \
  and\ \bibinfo {editor} {\bibfnamefont {B.}~\bibnamefont {Beltran}}\ (\bibinfo
  {year} {2008})\ pp.\ \bibinfo {pages} {19--50}\BibitemShut {NoStop}%
\bibitem [{\citenamefont {Sikivie}\ \emph {et~al.}(2014)\citenamefont
  {Sikivie}, \citenamefont {Sullivan},\ and\ \citenamefont
  {Tanner}}]{sikivie14}%
  \BibitemOpen
  \bibfield  {author} {\bibinfo {author} {\bibfnamefont {P}~\bibnamefont
  {Sikivie}}, \bibinfo {author} {\bibfnamefont {N}~\bibnamefont {Sullivan}}, \
  and\ \bibinfo {author} {\bibfnamefont {Dylan~B}\ \bibnamefont {Tanner}},\
  }\bibfield  {title} {\enquote {\bibinfo {title} {Proposal for axion dark
  matter detection using an l c circuit},}\ }\href {\doibase
  10.1103/PhysRevLett.112.131301} {\bibfield  {journal} {\bibinfo  {journal}
  {Phys. Rev. Lett.}\ }\textbf {\bibinfo {volume} {112}},\ \bibinfo {pages}
  {131301} (\bibinfo {year} {2014})}\BibitemShut {NoStop}%
\bibitem [{\citenamefont {Millar}\ \emph {et~al.}(2017)\citenamefont {Millar},
  \citenamefont {Redondo},\ and\ \citenamefont {Steffen}}]{millar17}%
  \BibitemOpen
  \bibfield  {author} {\bibinfo {author} {\bibfnamefont {Alexander~J}\
  \bibnamefont {Millar}}, \bibinfo {author} {\bibfnamefont {Javier}\
  \bibnamefont {Redondo}}, \ and\ \bibinfo {author} {\bibfnamefont {Frank~D}\
  \bibnamefont {Steffen}},\ }\bibfield  {title} {\enquote {\bibinfo {title}
  {Dielectric haloscopes: sensitivity to the axion dark matter velocity},}\
  }\href {\doibase 10.48550/arXiv.1707.04266} {\bibfield  {journal} {\bibinfo
  {journal} {J. Cosmol. Astropart. Phys.}\ }\textbf {\bibinfo {volume}
  {2017}},\ \bibinfo {pages} {006} (\bibinfo {year} {2017})}\BibitemShut
  {NoStop}%
\bibitem [{\citenamefont {Liu}\ \emph {et~al.}(2022)\citenamefont {Liu},
  \citenamefont {Dona}, \citenamefont {Hoshino}, \citenamefont {Knirck},
  \citenamefont {Kurinsky}, \citenamefont {Malaker}, \citenamefont {Miller},
  \citenamefont {Sonnenschein}, \citenamefont {Awida}, \citenamefont {Barry}
  \emph {et~al.}}]{liu22}%
  \BibitemOpen
  \bibfield  {author} {\bibinfo {author} {\bibfnamefont {Jesse}\ \bibnamefont
  {Liu}}, \bibinfo {author} {\bibfnamefont {Kristin}\ \bibnamefont {Dona}},
  \bibinfo {author} {\bibfnamefont {Gabe}\ \bibnamefont {Hoshino}}, \bibinfo
  {author} {\bibfnamefont {Stefan}\ \bibnamefont {Knirck}}, \bibinfo {author}
  {\bibfnamefont {Noah}\ \bibnamefont {Kurinsky}}, \bibinfo {author}
  {\bibfnamefont {Matthew}\ \bibnamefont {Malaker}}, \bibinfo {author}
  {\bibfnamefont {David~W}\ \bibnamefont {Miller}}, \bibinfo {author}
  {\bibfnamefont {Andrew}\ \bibnamefont {Sonnenschein}}, \bibinfo {author}
  {\bibfnamefont {Mohamed~H}\ \bibnamefont {Awida}}, \bibinfo {author}
  {\bibfnamefont {Peter~S}\ \bibnamefont {Barry}},  \emph {et~al.},\ }\bibfield
   {title} {\enquote {\bibinfo {title} {Broadband solenoidal haloscope for
  terahertz axion detection},}\ }\href {\doibase
  10.1103/PhysRevLett.128.131801} {\bibfield  {journal} {\bibinfo  {journal}
  {Phys. Rev. Lett.}\ }\textbf {\bibinfo {volume} {128}},\ \bibinfo {pages}
  {131--801} (\bibinfo {year} {2022})}\BibitemShut {NoStop}%
\bibitem [{\citenamefont {Li}\ \emph {et~al.}(1991)\citenamefont {Li},
  \citenamefont {Shi},\ and\ \citenamefont {Zhang}}]{li91}%
  \BibitemOpen
  \bibfield  {author} {\bibinfo {author} {\bibfnamefont {Xinzhou}\ \bibnamefont
  {Li}}, \bibinfo {author} {\bibfnamefont {Xin}\ \bibnamefont {Shi}}, \ and\
  \bibinfo {author} {\bibfnamefont {Jianzu}\ \bibnamefont {Zhang}},\ }\bibfield
   {title} {\enquote {\bibinfo {title} {Generalized riemann $\zeta$-function
  regularization and casimir energy for a piecewise uniform string},}\ }\href
  {\doibase 10.1103/PhysRevD.44.560} {\bibfield  {journal} {\bibinfo  {journal}
  {Phys. Rev. D}\ }\textbf {\bibinfo {volume} {44}},\ \bibinfo {pages} {560}
  (\bibinfo {year} {1991})}\BibitemShut {NoStop}%
\bibitem [{\citenamefont {Lawson}\ \emph {et~al.}(2019)\citenamefont {Lawson},
  \citenamefont {Millar}, \citenamefont {Pancaldi}, \citenamefont
  {Vitagliano},\ and\ \citenamefont {Wilczek}}]{lawson19}%
  \BibitemOpen
  \bibfield  {author} {\bibinfo {author} {\bibfnamefont {Matthew}\ \bibnamefont
  {Lawson}}, \bibinfo {author} {\bibfnamefont {Alexander~J}\ \bibnamefont
  {Millar}}, \bibinfo {author} {\bibfnamefont {Matteo}\ \bibnamefont
  {Pancaldi}}, \bibinfo {author} {\bibfnamefont {Edoardo}\ \bibnamefont
  {Vitagliano}}, \ and\ \bibinfo {author} {\bibfnamefont {Frank}\ \bibnamefont
  {Wilczek}},\ }\bibfield  {title} {\enquote {\bibinfo {title} {Tunable axion
  plasma haloscopes},}\ }\href {\doibase 10.1103/PhysRevLett.123.141802}
  {\bibfield  {journal} {\bibinfo  {journal} {Physical review letters}\
  }\textbf {\bibinfo {volume} {123}},\ \bibinfo {pages} {141802} (\bibinfo
  {year} {2019})}\BibitemShut {NoStop}%
\bibitem [{\citenamefont {Jiang}\ and\ \citenamefont
  {Wilczek}(2019)}]{qingdong19}%
  \BibitemOpen
  \bibfield  {author} {\bibinfo {author} {\bibfnamefont {Qing-Dong}\
  \bibnamefont {Jiang}}\ and\ \bibinfo {author} {\bibfnamefont {Frank}\
  \bibnamefont {Wilczek}},\ }\bibfield  {title} {\enquote {\bibinfo {title}
  {Chiral casimir forces: Repulsive, enhanced, tunable},}\ }\href {\doibase
  10.1103/PhysRevB.99.125403} {\bibfield  {journal} {\bibinfo  {journal}
  {Physical Review B}\ }\textbf {\bibinfo {volume} {99}},\ \bibinfo {pages}
  {125403} (\bibinfo {year} {2019})}\BibitemShut {NoStop}%
\bibitem [{\citenamefont {Sikivie}(2021)}]{sikivie03}%
  \BibitemOpen
  \bibfield  {author} {\bibinfo {author} {\bibfnamefont {Pierre}\ \bibnamefont
  {Sikivie}},\ }\bibfield  {title} {\enquote {\bibinfo {title} {Invisible axion
  search methods},}\ }\href {\doibase 10.1103/RevModPhys.93.015004} {\bibfield
  {journal} {\bibinfo  {journal} {Reviews of Modern Physics}\ }\textbf
  {\bibinfo {volume} {93}},\ \bibinfo {pages} {015004} (\bibinfo {year}
  {2021})}\BibitemShut {NoStop}%
\bibitem [{\citenamefont {McDonald}\ and\ \citenamefont
  {Ventura}(2020)}]{mcdonald20}%
  \BibitemOpen
  \bibfield  {author} {\bibinfo {author} {\bibfnamefont {Jamie~I}\ \bibnamefont
  {McDonald}}\ and\ \bibinfo {author} {\bibfnamefont {Lu{\'\i}s~B}\
  \bibnamefont {Ventura}},\ }\bibfield  {title} {\enquote {\bibinfo {title}
  {Optical properties of dynamical axion backgrounds},}\ }\href {\doibase
  10.1103/PhysRevD.101.123503} {\bibfield  {journal} {\bibinfo  {journal}
  {Physical Review D}\ }\textbf {\bibinfo {volume} {101}},\ \bibinfo {pages}
  {123503} (\bibinfo {year} {2020})}\BibitemShut {NoStop}%
\bibitem [{\citenamefont {Zyla}\ \emph {et~al.}(2020)\citenamefont {Zyla} \emph
  {et~al.}}]{zyla20}%
  \BibitemOpen
  \bibfield  {author} {\bibinfo {author} {\bibfnamefont {PA}~\bibnamefont
  {Zyla}} \emph {et~al.},\ }\bibfield  {title} {\enquote {\bibinfo {title} {79.
  baryon decay parameters},}\ }\href {\doibase 10.1093/ptep/ptaa104} {\bibfield
   {journal} {\bibinfo  {journal} {Prog. Theor. Exp. Phys}\ }\textbf {\bibinfo
  {volume} {2020}} (\bibinfo {year} {2020}),\ 10.1093/ptep/ptaa104}\BibitemShut
  {NoStop}%
\bibitem [{\citenamefont {Arza}\ \emph {et~al.}(2020)\citenamefont {Arza},
  \citenamefont {Schwetz},\ and\ \citenamefont {Todarello}}]{arza20}%
  \BibitemOpen
  \bibfield  {author} {\bibinfo {author} {\bibfnamefont {Ariel}\ \bibnamefont
  {Arza}}, \bibinfo {author} {\bibfnamefont {Thomas}\ \bibnamefont {Schwetz}},
  \ and\ \bibinfo {author} {\bibfnamefont {Elisa}\ \bibnamefont {Todarello}},\
  }\bibfield  {title} {\enquote {\bibinfo {title} {How to suppress exponential
  growth on the parametric resonance of photons in an axion background},}\
  }\href {\doibase 10.1088/1475-7516/2020/10/013} {\bibfield  {journal}
  {\bibinfo  {journal} {Journal of Cosmology and Astroparticle Physics}\
  }\textbf {\bibinfo {volume} {2020}},\ \bibinfo {pages} {013} (\bibinfo {year}
  {2020})}\BibitemShut {NoStop}%
\bibitem [{\citenamefont {Carenza}\ \emph {et~al.}(2020)\citenamefont
  {Carenza}, \citenamefont {Mirizzi},\ and\ \citenamefont {Sigl}}]{carenza20}%
  \BibitemOpen
  \bibfield  {author} {\bibinfo {author} {\bibfnamefont {Pierluca}\
  \bibnamefont {Carenza}}, \bibinfo {author} {\bibfnamefont {Alessandro}\
  \bibnamefont {Mirizzi}}, \ and\ \bibinfo {author} {\bibfnamefont
  {G{\"u}nter}\ \bibnamefont {Sigl}},\ }\bibfield  {title} {\enquote {\bibinfo
  {title} {Dynamical evolution of axion condensates under stimulated decays
  into photons},}\ }\href {\doibase 10.1103/PhysRevD.101.103016} {\bibfield
  {journal} {\bibinfo  {journal} {Physical Review D}\ }\textbf {\bibinfo
  {volume} {101}},\ \bibinfo {pages} {103016} (\bibinfo {year}
  {2020})}\BibitemShut {NoStop}%
\bibitem [{\citenamefont {Leroy}\ \emph {et~al.}(2020)\citenamefont {Leroy},
  \citenamefont {Chianese}, \citenamefont {Edwards},\ and\ \citenamefont
  {Weniger}}]{leroy20}%
  \BibitemOpen
  \bibfield  {author} {\bibinfo {author} {\bibfnamefont {Mika{\"e}l}\
  \bibnamefont {Leroy}}, \bibinfo {author} {\bibfnamefont {Marco}\ \bibnamefont
  {Chianese}}, \bibinfo {author} {\bibfnamefont {Thomas~DP}\ \bibnamefont
  {Edwards}}, \ and\ \bibinfo {author} {\bibfnamefont {Christoph}\ \bibnamefont
  {Weniger}},\ }\bibfield  {title} {\enquote {\bibinfo {title} {Radio signal of
  axion-photon conversion in neutron stars: A ray tracing analysis},}\ }\href
  {\doibase 10.1103/PhysRevD.101.123003} {\bibfield  {journal} {\bibinfo
  {journal} {Physical Review D}\ }\textbf {\bibinfo {volume} {101}},\ \bibinfo
  {pages} {123003} (\bibinfo {year} {2020})}\BibitemShut {NoStop}%
\bibitem [{\citenamefont {Ouellet}\ and\ \citenamefont
  {Bogorad}(2019)}]{oullet19}%
  \BibitemOpen
  \bibfield  {author} {\bibinfo {author} {\bibfnamefont {Jonathan}\
  \bibnamefont {Ouellet}}\ and\ \bibinfo {author} {\bibfnamefont {Zachary}\
  \bibnamefont {Bogorad}},\ }\bibfield  {title} {\enquote {\bibinfo {title}
  {Solutions to axion electrodynamics in various geometries},}\ }\href
  {\doibase 10.1103/PhysRevD.99.055010} {\bibfield  {journal} {\bibinfo
  {journal} {Physical Review D}\ }\textbf {\bibinfo {volume} {99}},\ \bibinfo
  {pages} {055010} (\bibinfo {year} {2019})}\BibitemShut {NoStop}%
\bibitem [{\citenamefont {Arza}\ and\ \citenamefont {Sikivie}(2019)}]{arza19}%
  \BibitemOpen
  \bibfield  {author} {\bibinfo {author} {\bibfnamefont {Ariel}\ \bibnamefont
  {Arza}}\ and\ \bibinfo {author} {\bibfnamefont {Pierre}\ \bibnamefont
  {Sikivie}},\ }\bibfield  {title} {\enquote {\bibinfo {title} {Production and
  detection of an axion dark matter echo},}\ }\href {\doibase
  10.1103/PhysRevLett.123.131804} {\bibfield  {journal} {\bibinfo  {journal}
  {Physical Review Letters}\ }\textbf {\bibinfo {volume} {123}},\ \bibinfo
  {pages} {131804} (\bibinfo {year} {2019})}\BibitemShut {NoStop}%
\bibitem [{\citenamefont {Qiu}\ \emph {et~al.}(2017)\citenamefont {Qiu},
  \citenamefont {Cao},\ and\ \citenamefont {Huang}}]{qiu17}%
  \BibitemOpen
  \bibfield  {author} {\bibinfo {author} {\bibfnamefont {Zebin}\ \bibnamefont
  {Qiu}}, \bibinfo {author} {\bibfnamefont {Gaoqing}\ \bibnamefont {Cao}}, \
  and\ \bibinfo {author} {\bibfnamefont {Xu-Guang}\ \bibnamefont {Huang}},\
  }\bibfield  {title} {\enquote {\bibinfo {title} {Electrodynamics of chiral
  matter},}\ }\href {\doibase 10.1103/PhysRevD.95.036002} {\bibfield  {journal}
  {\bibinfo  {journal} {Physical Review D}\ }\textbf {\bibinfo {volume} {95}},\
  \bibinfo {pages} {036002} (\bibinfo {year} {2017})}\BibitemShut {NoStop}%
\bibitem [{\citenamefont {Dror}\ \emph {et~al.}(2021)\citenamefont {Dror},
  \citenamefont {Murayama},\ and\ \citenamefont {Rodd}}]{dror21}%
  \BibitemOpen
  \bibfield  {author} {\bibinfo {author} {\bibfnamefont {Jeff~A}\ \bibnamefont
  {Dror}}, \bibinfo {author} {\bibfnamefont {Hitoshi}\ \bibnamefont
  {Murayama}}, \ and\ \bibinfo {author} {\bibfnamefont {Nicholas~L}\
  \bibnamefont {Rodd}},\ }\bibfield  {title} {\enquote {\bibinfo {title}
  {Cosmic axion background},}\ }\href {\doibase 10.1103/PhysRevD.103.115004}
  {\bibfield  {journal} {\bibinfo  {journal} {Physical Review D}\ }\textbf
  {\bibinfo {volume} {103}},\ \bibinfo {pages} {115004} (\bibinfo {year}
  {2021})}\BibitemShut {NoStop}%
\bibitem [{\citenamefont {Fukushima}\ \emph {et~al.}(2019)\citenamefont
  {Fukushima}, \citenamefont {Imaki},\ and\ \citenamefont {Qiu}}]{fukushima19}%
  \BibitemOpen
  \bibfield  {author} {\bibinfo {author} {\bibfnamefont {Kenji}\ \bibnamefont
  {Fukushima}}, \bibinfo {author} {\bibfnamefont {Shota}\ \bibnamefont
  {Imaki}}, \ and\ \bibinfo {author} {\bibfnamefont {Zebin}\ \bibnamefont
  {Qiu}},\ }\bibfield  {title} {\enquote {\bibinfo {title} {Anomalous casimir
  effect in axion electrodynamics},}\ }\href {\doibase
  10.1103/PhysRevD.100.045013} {\bibfield  {journal} {\bibinfo  {journal}
  {Physical Review D}\ }\textbf {\bibinfo {volume} {100}},\ \bibinfo {pages}
  {045013} (\bibinfo {year} {2019})}\BibitemShut {NoStop}%
\bibitem [{\citenamefont {Tobar}\ \emph {et~al.}(2019)\citenamefont {Tobar},
  \citenamefont {McAllister},\ and\ \citenamefont {Goryachev}}]{tobar19}%
  \BibitemOpen
  \bibfield  {author} {\bibinfo {author} {\bibfnamefont {Michael~E}\
  \bibnamefont {Tobar}}, \bibinfo {author} {\bibfnamefont {Ben~T}\ \bibnamefont
  {McAllister}}, \ and\ \bibinfo {author} {\bibfnamefont {Maxim}\ \bibnamefont
  {Goryachev}},\ }\bibfield  {title} {\enquote {\bibinfo {title} {Modified
  axion electrodynamics as impressed electromagnetic sources through
  oscillating background polarization and magnetization},}\ }\href {\doibase
  10.1016/j.dark.2019.100339} {\bibfield  {journal} {\bibinfo  {journal}
  {Physics of the Dark Universe}\ }\textbf {\bibinfo {volume} {26}},\ \bibinfo
  {pages} {100339} (\bibinfo {year} {2019})}\BibitemShut {NoStop}%
\bibitem [{\citenamefont {Bae}\ \emph {et~al.}(2023)\citenamefont {Bae},
  \citenamefont {Youn},\ and\ \citenamefont {Jeong}}]{bae22}%
  \BibitemOpen
  \bibfield  {author} {\bibinfo {author} {\bibfnamefont {Sungjae}\ \bibnamefont
  {Bae}}, \bibinfo {author} {\bibfnamefont {SungWoo}\ \bibnamefont {Youn}}, \
  and\ \bibinfo {author} {\bibfnamefont {Junu}\ \bibnamefont {Jeong}},\
  }\bibfield  {title} {\enquote {\bibinfo {title} {Tunable photonic crystal
  haloscope for high-mass axion searches},}\ }\href {\doibase
  10.1103/PhysRevD.107.015012} {\bibfield  {journal} {\bibinfo  {journal}
  {Physical Review D}\ }\textbf {\bibinfo {volume} {107}},\ \bibinfo {pages}
  {015012} (\bibinfo {year} {2023})}\BibitemShut {NoStop}%
\bibitem [{\citenamefont {Adshead}\ \emph {et~al.}(2020)\citenamefont
  {Adshead}, \citenamefont {Draper},\ and\ \citenamefont
  {Lillard}}]{adshead20}%
  \BibitemOpen
  \bibfield  {author} {\bibinfo {author} {\bibfnamefont {Peter}\ \bibnamefont
  {Adshead}}, \bibinfo {author} {\bibfnamefont {Patrick}\ \bibnamefont
  {Draper}}, \ and\ \bibinfo {author} {\bibfnamefont {Benjamin}\ \bibnamefont
  {Lillard}},\ }\bibfield  {title} {\enquote {\bibinfo {title} {Time-domain
  properties of electromagnetic signals in a dynamical axion background},}\
  }\href {\doibase 10.1103/PhysRevD.102.123011} {\bibfield  {journal} {\bibinfo
   {journal} {Physical Review D}\ }\textbf {\bibinfo {volume} {102}},\ \bibinfo
  {pages} {123011} (\bibinfo {year} {2020})}\BibitemShut {NoStop}%
\bibitem [{\citenamefont {Tobar}\ \emph {et~al.}(2022)\citenamefont {Tobar},
  \citenamefont {McAllister},\ and\ \citenamefont {Goryachev}}]{tobar22}%
  \BibitemOpen
  \bibfield  {author} {\bibinfo {author} {\bibfnamefont {Michael~E}\
  \bibnamefont {Tobar}}, \bibinfo {author} {\bibfnamefont {Ben~T}\ \bibnamefont
  {McAllister}}, \ and\ \bibinfo {author} {\bibfnamefont {Maxim}\ \bibnamefont
  {Goryachev}},\ }\bibfield  {title} {\enquote {\bibinfo {title} {Poynting
  vector controversy in axion modified electrodynamics},}\ }\href {\doibase
  10.1103/PhysRevD.105.045009} {\bibfield  {journal} {\bibinfo  {journal}
  {Physical Review D}\ }\textbf {\bibinfo {volume} {105}},\ \bibinfo {pages}
  {045009} (\bibinfo {year} {2022})}\BibitemShut {NoStop}%
\bibitem [{\citenamefont {DeRocco}\ and\ \citenamefont
  {Hook}(2018)}]{derocco18}%
  \BibitemOpen
  \bibfield  {author} {\bibinfo {author} {\bibfnamefont {William}\ \bibnamefont
  {DeRocco}}\ and\ \bibinfo {author} {\bibfnamefont {Anson}\ \bibnamefont
  {Hook}},\ }\bibfield  {title} {\enquote {\bibinfo {title} {Axion
  interferometry},}\ }\href {\doibase 10.1103/PhysRevD.98.035021} {\bibfield
  {journal} {\bibinfo  {journal} {Physical Review D}\ }\textbf {\bibinfo
  {volume} {98}},\ \bibinfo {pages} {035021} (\bibinfo {year}
  {2018})}\BibitemShut {NoStop}%
\bibitem [{\citenamefont {Brevik}\ and\ \citenamefont
  {Chaichian}(2022{\natexlab{a}})}]{brevik22}%
  \BibitemOpen
  \bibfield  {author} {\bibinfo {author} {\bibfnamefont {Iver~H}\ \bibnamefont
  {Brevik}}\ and\ \bibinfo {author} {\bibfnamefont {Moshe~M}\ \bibnamefont
  {Chaichian}},\ }\bibfield  {title} {\enquote {\bibinfo {title} {Axion
  electrodynamics: Energy--momentum tensor and possibilities for experimental
  tests},}\ }\href {\doibase 10.1142/S0217751X22501512} {\bibfield  {journal}
  {\bibinfo  {journal} {International Journal of Modern Physics A}\ }\textbf
  {\bibinfo {volume} {37}},\ \bibinfo {pages} {2250151} (\bibinfo {year}
  {2022}{\natexlab{a}})}\BibitemShut {NoStop}%
\bibitem [{\citenamefont {Brevik}\ and\ \citenamefont
  {Chaichian}(2022{\natexlab{b}})}]{brevik22a}%
  \BibitemOpen
  \bibfield  {author} {\bibinfo {author} {\bibfnamefont {Iver~H}\ \bibnamefont
  {Brevik}}\ and\ \bibinfo {author} {\bibfnamefont {Moshe~M}\ \bibnamefont
  {Chaichian}},\ }\bibfield  {title} {\enquote {\bibinfo {title} {Electric
  current and heat production by a neutral carrier: an effect of the axion},}\
  }\href {\doibase 10.1140/epjc/s10052-022-10150-1} {\bibfield  {journal}
  {\bibinfo  {journal} {The European Physical Journal C}\ }\textbf {\bibinfo
  {volume} {82}},\ \bibinfo {pages} {202} (\bibinfo {year}
  {2022}{\natexlab{b}})}\BibitemShut {NoStop}%
\bibitem [{\citenamefont {Brevik}\ \emph
  {et~al.}(2023{\natexlab{a}})\citenamefont {Brevik}, \citenamefont {Favitta},\
  and\ \citenamefont {Chaichian}}]{brevik23}%
  \BibitemOpen
  \bibfield  {author} {\bibinfo {author} {\bibfnamefont {Iver}\ \bibnamefont
  {Brevik}}, \bibinfo {author} {\bibfnamefont {Amedeo~M}\ \bibnamefont
  {Favitta}}, \ and\ \bibinfo {author} {\bibfnamefont {Masud}\ \bibnamefont
  {Chaichian}},\ }\bibfield  {title} {\enquote {\bibinfo {title} {Axionic and
  nonaxionic electrodynamics in plane and circular geometry},}\ }\href
  {\doibase 10.1103/PhysRevD.107.043522} {\bibfield  {journal} {\bibinfo
  {journal} {Physical Review D}\ }\textbf {\bibinfo {volume} {107}},\ \bibinfo
  {pages} {043522} (\bibinfo {year} {2023}{\natexlab{a}})}\BibitemShut
  {NoStop}%
\bibitem [{\citenamefont {Brevik}\ \emph
  {et~al.}(2023{\natexlab{b}})\citenamefont {Brevik}, \citenamefont
  {Chaichian},\ and\ \citenamefont {Favitta}}]{favitta23}%
  \BibitemOpen
  \bibfield  {author} {\bibinfo {author} {\bibfnamefont {Iver}\ \bibnamefont
  {Brevik}}, \bibinfo {author} {\bibfnamefont {Masud}\ \bibnamefont
  {Chaichian}}, \ and\ \bibinfo {author} {\bibfnamefont {Amedeo~M}\
  \bibnamefont {Favitta}},\ }\bibfield  {title} {\enquote {\bibinfo {title} {On
  the axion electrodynamics in a two-dimensional slab and the casimir
  effect},}\ }\href {\doibase 10.48550/arXiv.2310.05575} {\bibfield  {journal}
  {\bibinfo  {journal} {arXiv preprint arXiv:2310.05575}\ } (\bibinfo {year}
  {2023}{\natexlab{b}}),\ 10.48550/arXiv.2310.05575}\BibitemShut {NoStop}%
\bibitem [{\citenamefont {Qi}\ and\ \citenamefont
  {Zhang}(2011)}]{qi2011topological}%
  \BibitemOpen
  \bibfield  {author} {\bibinfo {author} {\bibfnamefont {Xiao-Liang}\
  \bibnamefont {Qi}}\ and\ \bibinfo {author} {\bibfnamefont {Shou-Cheng}\
  \bibnamefont {Zhang}},\ }\bibfield  {title} {\enquote {\bibinfo {title}
  {Topological insulators and superconductors},}\ }\href {\doibase
  10.1103/RevModPhys.83.1057} {\bibfield  {journal} {\bibinfo  {journal}
  {Reviews of Modern Physics}\ }\textbf {\bibinfo {volume} {83}},\ \bibinfo
  {pages} {1057} (\bibinfo {year} {2011})}\BibitemShut {NoStop}%
\bibitem [{\citenamefont {Qi}\ \emph {et~al.}(2008)\citenamefont {Qi},
  \citenamefont {Hughes},\ and\ \citenamefont {Zhang}}]{qi2008topological}%
  \BibitemOpen
  \bibfield  {author} {\bibinfo {author} {\bibfnamefont {Xiao-Liang}\
  \bibnamefont {Qi}}, \bibinfo {author} {\bibfnamefont {Taylor~L}\ \bibnamefont
  {Hughes}}, \ and\ \bibinfo {author} {\bibfnamefont {Shou-Cheng}\ \bibnamefont
  {Zhang}},\ }\bibfield  {title} {\enquote {\bibinfo {title} {Topological field
  theory of time-reversal invariant insulators},}\ }\href {\doibase
  10.1103/PhysRevB.78.195424} {\bibfield  {journal} {\bibinfo  {journal}
  {Physical Review B}\ }\textbf {\bibinfo {volume} {78}},\ \bibinfo {pages}
  {195424} (\bibinfo {year} {2008})}\BibitemShut {NoStop}%
\bibitem [{\citenamefont {Peccei}\ and\ \citenamefont
  {Quinn}(1977{\natexlab{b}})}]{peccei1977cp}%
  \BibitemOpen
  \bibfield  {author} {\bibinfo {author} {\bibfnamefont {Roberto~D}\
  \bibnamefont {Peccei}}\ and\ \bibinfo {author} {\bibfnamefont {Helen~R}\
  \bibnamefont {Quinn}},\ }\bibfield  {title} {\enquote {\bibinfo {title} {Cp
  conservation in the presence of pseudoparticles},}\ }\href {\doibase
  10.1103/PhysRevLett.38.1440} {\bibfield  {journal} {\bibinfo  {journal}
  {Physical Review Letters}\ }\textbf {\bibinfo {volume} {38}},\ \bibinfo
  {pages} {1440} (\bibinfo {year} {1977}{\natexlab{b}})}\BibitemShut {NoStop}%
\bibitem [{\citenamefont {Peccei}\ and\ \citenamefont
  {Quinn}(1977{\natexlab{c}})}]{peccei1977constraints}%
  \BibitemOpen
  \bibfield  {author} {\bibinfo {author} {\bibfnamefont {Roberto~D}\
  \bibnamefont {Peccei}}\ and\ \bibinfo {author} {\bibfnamefont {Helen~R}\
  \bibnamefont {Quinn}},\ }\bibfield  {title} {\enquote {\bibinfo {title}
  {Constraints imposed by cp conservation in the presence of
  pseudoparticles},}\ }\href {\doibase 10.1103/PhysRevD.16.1791} {\bibfield
  {journal} {\bibinfo  {journal} {Physical Review D}\ }\textbf {\bibinfo
  {volume} {16}},\ \bibinfo {pages} {1791} (\bibinfo {year}
  {1977}{\natexlab{c}})}\BibitemShut {NoStop}%
\bibitem [{\citenamefont {Lu}(2021)}]{lu2021casimir}%
  \BibitemOpen
  \bibfield  {author} {\bibinfo {author} {\bibfnamefont {Bing-Sui}\
  \bibnamefont {Lu}},\ }\bibfield  {title} {\enquote {\bibinfo {title} {The
  casimir effect in topological matter},}\ }\href {\doibase
  10.3390/universe7070237} {\bibfield  {journal} {\bibinfo  {journal}
  {Universe}\ }\textbf {\bibinfo {volume} {7}},\ \bibinfo {pages} {237}
  (\bibinfo {year} {2021})}\BibitemShut {NoStop}%
\bibitem [{\citenamefont {Mart{\'\i}n-Ruiz}\ \emph {et~al.}(2019)\citenamefont
  {Mart{\'\i}n-Ruiz}, \citenamefont {Cambiaso},\ and\ \citenamefont
  {Urrutia}}]{martin2019magnetoelectric}%
  \BibitemOpen
  \bibfield  {author} {\bibinfo {author} {\bibfnamefont {A}~\bibnamefont
  {Mart{\'\i}n-Ruiz}}, \bibinfo {author} {\bibfnamefont {M}~\bibnamefont
  {Cambiaso}}, \ and\ \bibinfo {author} {\bibfnamefont {LF}~\bibnamefont
  {Urrutia}},\ }\bibfield  {title} {\enquote {\bibinfo {title} {The
  magnetoelectric coupling in electrodynamics},}\ }\href {\doibase
  10.1142/S0217751X19410021} {\bibfield  {journal} {\bibinfo  {journal}
  {International Journal of Modern Physics A}\ }\textbf {\bibinfo {volume}
  {34}},\ \bibinfo {pages} {1941002} (\bibinfo {year} {2019})}\BibitemShut
  {NoStop}%
\bibitem [{\citenamefont {Nogueira}\ and\ \citenamefont {van~den
  Brink}(2022)}]{nogueira2022absence}%
  \BibitemOpen
  \bibfield  {author} {\bibinfo {author} {\bibfnamefont {Flavio~S}\
  \bibnamefont {Nogueira}}\ and\ \bibinfo {author} {\bibfnamefont {Jeroen}\
  \bibnamefont {van~den Brink}},\ }\bibfield  {title} {\enquote {\bibinfo
  {title} {Absence of induced magnetic monopoles in maxwellian
  magnetoelectrics},}\ }\href {\doibase 10.1103/PhysRevResearch.4.013074}
  {\bibfield  {journal} {\bibinfo  {journal} {Physical Review Research}\
  }\textbf {\bibinfo {volume} {4}},\ \bibinfo {pages} {013074} (\bibinfo {year}
  {2022})}\BibitemShut {NoStop}%
\bibitem [{\citenamefont {Woods}\ \emph {et~al.}(2020)\citenamefont {Woods},
  \citenamefont {Kr{\"u}ger},\ and\ \citenamefont
  {Dodonov}}]{woods2020perspective}%
  \BibitemOpen
  \bibfield  {author} {\bibinfo {author} {\bibfnamefont {Lilia~M}\ \bibnamefont
  {Woods}}, \bibinfo {author} {\bibfnamefont {Matthias}\ \bibnamefont
  {Kr{\"u}ger}}, \ and\ \bibinfo {author} {\bibfnamefont {Victor~V}\
  \bibnamefont {Dodonov}},\ }\bibfield  {title} {\enquote {\bibinfo {title}
  {Perspective on some recent and future developments in casimir
  interactions},}\ }\href {\doibase 10.3390/app11010293} {\bibfield  {journal}
  {\bibinfo  {journal} {Applied Sciences}\ }\textbf {\bibinfo {volume} {11}},\
  \bibinfo {pages} {293} (\bibinfo {year} {2020})}\BibitemShut {NoStop}%
\bibitem [{\citenamefont {To}\ \emph {et~al.}(2022)\citenamefont {To},
  \citenamefont {Wang}, \citenamefont {Ho}, \citenamefont {Hu}, \citenamefont
  {Acuna}, \citenamefont {Liu}, \citenamefont {Bryant}, \citenamefont
  {Janotti}, \citenamefont {Zide}, \citenamefont {Law} \emph
  {et~al.}}]{to2022strong}%
  \BibitemOpen
  \bibfield  {author} {\bibinfo {author} {\bibfnamefont {D~Quang}\ \bibnamefont
  {To}}, \bibinfo {author} {\bibfnamefont {Zhengtianye}\ \bibnamefont {Wang}},
  \bibinfo {author} {\bibfnamefont {Dai~Q}\ \bibnamefont {Ho}}, \bibinfo
  {author} {\bibfnamefont {Ruiqi}\ \bibnamefont {Hu}}, \bibinfo {author}
  {\bibfnamefont {Wilder}\ \bibnamefont {Acuna}}, \bibinfo {author}
  {\bibfnamefont {Yongchen}\ \bibnamefont {Liu}}, \bibinfo {author}
  {\bibfnamefont {Garnett~W}\ \bibnamefont {Bryant}}, \bibinfo {author}
  {\bibfnamefont {Anderson}\ \bibnamefont {Janotti}}, \bibinfo {author}
  {\bibfnamefont {Joshua~MO}\ \bibnamefont {Zide}}, \bibinfo {author}
  {\bibfnamefont {Stephanie}\ \bibnamefont {Law}},  \emph {et~al.},\ }\bibfield
   {title} {\enquote {\bibinfo {title} {Strong coupling between a topological
  insulator and a iii-v heterostructure at terahertz frequency},}\ }\href
  {\doibase 10.1103/PhysRevMaterials.6.035201} {\bibfield  {journal} {\bibinfo
  {journal} {Physical Review Materials}\ }\textbf {\bibinfo {volume} {6}},\
  \bibinfo {pages} {035201} (\bibinfo {year} {2022})}\BibitemShut {NoStop}%
\bibitem [{\citenamefont {Kargarian}\ \emph {et~al.}(2015)\citenamefont
  {Kargarian}, \citenamefont {Randeria},\ and\ \citenamefont
  {Trivedi}}]{kargarian2015theory}%
  \BibitemOpen
  \bibfield  {author} {\bibinfo {author} {\bibfnamefont {Mehdi}\ \bibnamefont
  {Kargarian}}, \bibinfo {author} {\bibfnamefont {Mohit}\ \bibnamefont
  {Randeria}}, \ and\ \bibinfo {author} {\bibfnamefont {Nandini}\ \bibnamefont
  {Trivedi}},\ }\bibfield  {title} {\enquote {\bibinfo {title} {Theory of kerr
  and faraday rotations and linear dichroism in topological weyl semimetals},}\
  }\href {\doibase 10.1038/srep12683} {\bibfield  {journal} {\bibinfo
  {journal} {Scientific reports}\ }\textbf {\bibinfo {volume} {5}},\ \bibinfo
  {pages} {12683} (\bibinfo {year} {2015})}\BibitemShut {NoStop}%
\bibitem [{\citenamefont {Wilson}\ \emph {et~al.}(2015)\citenamefont {Wilson},
  \citenamefont {Allocca},\ and\ \citenamefont
  {Galitski}}]{wilson2015repulsive}%
  \BibitemOpen
  \bibfield  {author} {\bibinfo {author} {\bibfnamefont {Justin~H}\
  \bibnamefont {Wilson}}, \bibinfo {author} {\bibfnamefont {Andrew~A}\
  \bibnamefont {Allocca}}, \ and\ \bibinfo {author} {\bibfnamefont {Victor}\
  \bibnamefont {Galitski}},\ }\bibfield  {title} {\enquote {\bibinfo {title}
  {Repulsive casimir force between weyl semimetals},}\ }\href {\doibase
  10.1103/PhysRevB.91.235115} {\bibfield  {journal} {\bibinfo  {journal}
  {Physical Review B}\ }\textbf {\bibinfo {volume} {91}},\ \bibinfo {pages}
  {235115} (\bibinfo {year} {2015})}\BibitemShut {NoStop}%
\bibitem [{\citenamefont {Mart{\'\i}n-Ruiz}\ \emph {et~al.}(2016)\citenamefont
  {Mart{\'\i}n-Ruiz}, \citenamefont {Cambiaso},\ and\ \citenamefont
  {Urrutia}}]{martin2016green}%
  \BibitemOpen
  \bibfield  {author} {\bibinfo {author} {\bibfnamefont {A}~\bibnamefont
  {Mart{\'\i}n-Ruiz}}, \bibinfo {author} {\bibfnamefont {M}~\bibnamefont
  {Cambiaso}}, \ and\ \bibinfo {author} {\bibfnamefont {LF}~\bibnamefont
  {Urrutia}},\ }\bibfield  {title} {\enquote {\bibinfo {title} {A green's
  function approach to the casimir effect on topological insulators with planar
  symmetry},}\ }\href {\doibase 10.1209/0295-5075/113/60005} {\bibfield
  {journal} {\bibinfo  {journal} {Europhysics Letters}\ }\textbf {\bibinfo
  {volume} {113}},\ \bibinfo {pages} {60005} (\bibinfo {year}
  {2016})}\BibitemShut {NoStop}%
\bibitem [{\citenamefont {Woods}\ \emph {et~al.}(2016)\citenamefont {Woods},
  \citenamefont {Dalvit}, \citenamefont {Tkatchenko}, \citenamefont
  {Rodriguez-Lopez}, \citenamefont {Rodriguez},\ and\ \citenamefont
  {Podgornik}}]{woods2016materials}%
  \BibitemOpen
  \bibfield  {author} {\bibinfo {author} {\bibfnamefont {LM}~\bibnamefont
  {Woods}}, \bibinfo {author} {\bibfnamefont {Diego Alejandro~Roberto}\
  \bibnamefont {Dalvit}}, \bibinfo {author} {\bibfnamefont {Alexandre}\
  \bibnamefont {Tkatchenko}}, \bibinfo {author} {\bibfnamefont {P}~\bibnamefont
  {Rodriguez-Lopez}}, \bibinfo {author} {\bibfnamefont {Alejandro~W}\
  \bibnamefont {Rodriguez}}, \ and\ \bibinfo {author} {\bibfnamefont
  {R}~\bibnamefont {Podgornik}},\ }\bibfield  {title} {\enquote {\bibinfo
  {title} {Materials perspective on casimir and van der waals interactions},}\
  }\href {\doibase 10.1103/RevModPhys.88.045003} {\bibfield  {journal}
  {\bibinfo  {journal} {Reviews of Modern Physics}\ }\textbf {\bibinfo {volume}
  {88}},\ \bibinfo {pages} {045003} (\bibinfo {year} {2016})}\BibitemShut
  {NoStop}%
\bibitem [{\citenamefont {Khusnutdinov}\ and\ \citenamefont
  {Woods}(2019)}]{khusnutdinov2019casimir}%
  \BibitemOpen
  \bibfield  {author} {\bibinfo {author} {\bibfnamefont {N}~\bibnamefont
  {Khusnutdinov}}\ and\ \bibinfo {author} {\bibfnamefont {LM}~\bibnamefont
  {Woods}},\ }\bibfield  {title} {\enquote {\bibinfo {title} {Casimir effects
  in 2d dirac materials (scientific summary)},}\ }\href {\doibase
  10.1134/S0021364019150013} {\bibfield  {journal} {\bibinfo  {journal} {JETP
  Letters}\ }\textbf {\bibinfo {volume} {110}},\ \bibinfo {pages} {183--192}
  (\bibinfo {year} {2019})}\BibitemShut {NoStop}%
\bibitem [{\citenamefont {Brevik}(2021)}]{brevik21}%
  \BibitemOpen
  \bibfield  {author} {\bibinfo {author} {\bibfnamefont {Iver}\ \bibnamefont
  {Brevik}},\ }\bibfield  {title} {\enquote {\bibinfo {title} {Axion
  electrodynamics and the axionic casimir effect},}\ }\href
  {https://www.mdpi.com/2218-1997/7/5/133} {\bibfield  {journal} {\bibinfo
  {journal} {Universe}\ }\textbf {\bibinfo {volume} {7}} (\bibinfo {year}
  {2021})}\BibitemShut {NoStop}%
\bibitem [{Note1()}]{Note1}%
  \BibitemOpen
  \bibinfo {note} {Note that the present definition of $\beta $ differs from
  that in Ref.~\cite {brevik21}.}\BibitemShut {Stop}%
\end{thebibliography}%

\end{document}